\documentclass{article}

\usepackage{authblk}
\usepackage[utf8]{inputenc}
\usepackage{url}
\usepackage{booktabs}
\usepackage{amsfonts}
\usepackage{microtype}
\usepackage{graphicx}
\usepackage{float}
\usepackage{miller}
\usepackage{url}
\usepackage{siunitx}
\DeclareSIUnit\atom{at}
\DeclareSIUnit\angstrom{\text{Å}}
\DeclareSIUnit\electrons{\text{e}}
\usepackage{xcolor}
\usepackage{graphicx}
\usepackage[numbers]{natbib}
\usepackage{doi}
\usepackage{hyperref}

\newcommand{\Fcrit}{\ensuremath{F_\text{c}}}
\newcommand{\Fp}{\ensuremath{F_\text{P}}}
\newcommand{\FpP}{\ensuremath{F_{\text{P},95}}}

\hypersetup{
  pdftitle={The origin of jerky dislocation motion in high-entropy alloys},
  pdfauthor={Daniel Utt, Subin Lee, Yaolong Xing, Hyejin Jeong, Alexander Stukowski, Sang Ho Oh, Gerhard Dehm, Karsten Albe},
}

\title{The origin of jerky dislocation motion in high-entropy alloys}

\author[1]{Daniel Utt}
\author[2,3,4]{Subin Lee}
\author[5]{Yaolong Xing}
\author[5]{Hyejin Jeong}
\author[1]{Alexander Stukowski}
\author[5,6]{Sang Ho Oh}
\author[2]{Gerhard Dehm}
\author[1]{Karsten Albe}
\affil[1]{Fachgebiet Materialmodellierung, Institut für Materialwissenschaft, Technische Universität Darmstadt, Darmstadt, Germany}
\affil[2]{Structure and Nano-/Micromechanics of Materials, Max-Planck-Institut für Eisenforschung GmbH, Düsseldorf, Germany}
\affil[3]{Center for Integrated Nanostructure Physics, Institute for Basic Science, Suwon, Republic of Korea}
\affil[4]{Institute for Applied Materials, Karlsruhe Institute of Technology,
Eggenstein-Leopoldshafen, Germany}
\affil[5]{Department of Energy Science, Sungkyunkwan University, Suwon, Republic of Korea}
\affil[6]{Department of Energy Engineering, KENTECH Institute for Energy Materials and Devices, Korea Institute of Energy Technology (KENTECH), Naju, Republic of Korea}
\date{}

\begin{document}

\maketitle

\begin{abstract}
  Dislocations in single-phase concentrated random alloys, including high- entropy alloys (HEAs), repeatedly encounter pinning during glide, resulting in jerky dislocation motion. While solute-dislocation interaction is well understood in conventional alloys, the origin of individual pinning points in concentrated random alloys is a matter of debate. In this work, we investigate the origin of dislocation pinning in the CoCrFeMnNi HEA. In- situ transmission electron microscopy studies reveal wavy dislocation lines and a jagged glide motion under external loading, even though no segregation or clustering is found around Shockley partial dislocations. Atomistic simulations reproduce the jerky dislocation motion and link the repeated pinning to local fluctuations in the Peierls friction. We demonstrate that the density of high local Peierls friction is proportional to the critical stress required for dislocation glide and the dislocation mobility.
\end{abstract}

\section{Introduction}

High- and medium-entropy alloys (HEAs and MEAs) are a new class of metallic materials that contain multiple elements at high concentrations and form solid solutions\cite{George2019} in contrast to conventional alloys that typically consist of a single principal element with low concentrations of secondary elements. Since entropy is not always the decisive design parameter, they are a subclass of concentrated random alloys. Several HEAs outrival the mechanical properties of conventional alloys, some possess exclusive property combinations such as high strength and high ductility—even down to cryogenic temperatures\cite{Wu2014a,Gludovatz2014,Li2016a,George2020,Raabe2015}. The origin of the high strength of HEAs is key for materials design but is still controversially discussed in the community\cite{Li2019d,Yin2020a,Zhang2020a,Nohring2020,Ma2020}. There are, however, four different origins explored in the literature to date:
\begin{enumerate}
\item Short-range ordering (SRO), i.e., a preferred ordering in the atomic-sized neighborhood, is well known to lead to a strength increase in conventional alloys\cite{Gerold1989}. This strengthening mechanism has been reported for the CoCrNi, a subsystem of the Cantor (CoCrFeMnNi) alloy\cite{Li2019d,Zhang2020a,Antillon2020}. Since trains of dislocations traveling on the same plane were observed in this system, it is argued that the first dislocation gliding on a plane destroys SRO within that plane and facilitates the glide of subsequent dislocations\cite{Otto2013}. Yin and Curtin\cite{Yin2020a}, on the other hand, showed that the extraordinary strength of CoCrNi\cite{Wu2014a} can also be explained based on solid solution strengthening without resorting to SRO.
\item Indeed, dislocation pinning and the resulting strengthening are also observed in ideally random alloys\cite{Wu2014a}. Li et al.\cite{Li2019d} reported that even in a perfectly random CoCrNi alloy some preferential atomic arrangements, in the case of Co and Cr, exist, and breaking these randomly appearing favorable bonds requires additional energy.
\item The unexpected wide range of stacking fault (SF) energies observed for HEAs could provide an alternative explanation for the pinning experienced by dislocations\cite{Zeng2019,Smith2016}. The resulting wavy dislocation lines have been observed for different HEAs in transmission electron microscopy (TEM) experiments\cite{Smith2016,Lee2020,Laplanche2016} and atomistic simulations\cite{Li2019d,Rao2017a,Pasianot2020}. Li et al.\cite{Li2019d} argued that the wide range of stacking fault energy (SFE) causes peaks in the energy landscape where lattice friction becomes enhanced locally. Therefore, dislocation segments need to unzip from the local barriers while gliding. Such a mechanism would only exist in HEAs, but not in conventional dilute solid solutions as their local fluctuations in the SF energy remain comparatively small.
\item Lastly, the solid solution strengthening model developed by Varvenne et al.\cite{Varvenne2016a} predicts the flow stresses in an HEA solely based on misfit volumes and elastic properties. The dominating mechanism of solid solution strengthening in HEAs is intricate as it becomes impossible to define distinct matrix and solute atoms. A Labusch-type weak interaction\cite{Labusch1970} between solute atoms and dislocations is assumed in Varvenne's model. Here, the dislocation line interacts with the whole field of solutes within its elastic interaction range at the same time. This is opposite to the strong-pinning model from Fleischer\cite{Fleischer1963} where each solute corresponds to a site where the dislocation line is pinned.
\end{enumerate}

In this study, we investigate the repeated local pinning along subsequent dislocations in the face-centered cubic (FCC) CoCrFeMnNi HEA by combining in- situ TEM deformation, atomic resolution analytical scanning transmission electron microscopy (STEM), and atomistic simulations. In-situ TEM deformation tests reveal wavy dislocation lines and a jagged dislocation motion under an applied tensile strain, although no indication of SRO is found by atomic resolution STEM energy-dispersive X-ray spectroscopy (EDS). Atomistic computer simulations reproduce the jerky dislocation motion with repeated pinning in random samples without SRO. The underlying mechanism is directly linked to local fluctuations in Peierls (force/friction) barriers, which lead to dislocation pinning points in concentrated random alloys. We further show their relation to the critical stress needed to move a dislocation line and the resulting mobility. While not a full solid solution strengthening model, the atomic fluctuations in frictional forces provide predictive design guidelines to tailor the dislocation pinning in HEAs.

\section{Methods}
\subsection{Fabrication of CoCrFeMnNi single crystal}

A single crystal CoCrFeMnNi HEA was produced by using the Czochralski method from an equiatomic melt of high-purity elements. The composition of Fe, Co, Cr, Mn, and Ni is 24, 24, 20, 12, 20\,\si{\atom\percent}, respectively, measured by both STEM-EDS and atom probe tomography\cite{Lee2020,Raghavan2017}. In situ TEM straining samples were prepared by focused ion beam (FIB) site-specific lift-out method\cite{Li2018d} that utilizes a 30 and \SI{5}{\kilo\volt} Ga ion beam (JIB-4610F, JEOL). A thin lamella with a size of several micrometers was glued on a custom-made Cu grid by Pt deposition.

\subsection{In situ TEM tensile experiments}

In situ tensile straining tests were carried out in an aberration-corrected TEM operated at \SI{300}{\kilo\volt} (JEM-ARM300CF, JEOL) equipped with a straining holder (Model 654, Gatan). In situ TEM movies were recorded with a high-speed camera at a frame rate of 50 frames per second and 2k resolution (Oneview, Gatan). The displacement of the whole \(\sim \SI{9}{\milli\meter}\) large sample support was controlled by a step motor whose resolution was \SI{1}{\micro\meter}. The straining was interrupted intermittently to stabilize the stage and observe dislocation glide motion. No quantitative load-displacement data can be obtained during the deformation.

The contrast of dislocations in TEM images and videos is enhanced by image processing. Firstly, the background noise from the FIB damage is removed by using direct subtraction of the last frame after aligning all frames using the cross-correlation method. Then, a Gaussian filter is used to blur the remaining noise. To extract dislocation positions as shown in Fig.~\ref{fig1}b, a ridge detection filter is used after applying an additional Gaussian filter.

\subsection{STEM EDS experiments}

The atomic-resolution STEM EDS chemical mapping was carried out on a JEM-ARM 200F (JEOL) equipped with a spherical aberration corrector (ASCOR, CEOS) and energy dispersive X-ray spectrometer (JED-EDS, JEOL). Multi-frame EDS maps were acquired for a half-hour by using dual-type EDS detectors (the effective X-ray detection area of a \SI{100}{\milli\meter\squared} for each detector) with a large effective solid angle (\(\sim\SI{1.2}{sr}\)) and a highly focused electron probe (\(\sim\SI{1.2}{\angstrom}\)) at the electron dose rate of \SI{6.3e9}{\electrons\per\nano\meter\squared\per\second}. The resulting elemental maps were obtained by the multiple frame summation up to \num{2300} frames with \(256 \times 256\) pixels resolution and the acquisition time of \SI{10}{\micro\second} per pixel (\(\sim\SI{20}{\minute}\) in maximum as a total acquisition time). Multi-frame EDS maps were averaged by correcting the frame-to-frame displacement caused by sample drift during acquisition. The averaged X-ray count of each element was converted to the composition by calibrating the k-factor, which accounts for the different X-ray yields of each element. The background noise floor in each composition map was suppressed by averaging every neighboring \num{5} pixels.

\subsection{Atomistic simulations}

  The equimolar CoCrFeMnNi HEA and its subsystems were all simulated based on the 2NN MEAM\cite{Lee2000} interatomic potential by Choi et al.\cite{Choi2018}, the CoCrCuFeNi\cite{Pasianot2020,Farkas2018} and the Co\(_{30}\)Fe\(_{17}\)Ni\(_{37}\)Ti\(_{17}\)\cite{Rao2017a,Zhou2004} HEA are based on the Farkas and Caro\cite{Farkas2018} or Zhou et al.\cite{Zhou2004} EAM potentials\cite{Daw1983,Daw1984}, respectively, to exclude erroneous correlations from a given parametrization. All simulations were run in \textsc{LAMMPS}\cite{Thompson2022}. Samples were prepared using \textsc{ATOMSK}\cite{Hirel2015}, and post-processing was based on algorithms implemented in \textsc{OVITO}\cite{Stukowski2010}, and accelerated by \textsc{PARALLEL}\cite{Tange2011a}.

\subsection{Virtual TEM lamella}

The large-scale sample shown in Fig.~\ref{fig1} measurements had initial dimensions of \(101 \times 152 \times 75\)\,\si{\nano\meter\cubed} of pristine FCC lattice oriented \hkl[-1-2-3], \hkl[1-2-1], \hkl[412] along \(x\), \(y\), and \(z\), respectively. This gave approximately \(10^8\) lattice sites, filled with either CoCrFeMnNi or CoNi alloy. A surface notch with an inclination of \SI{28.13}{\degree} was cut into the \(y-z\) plane. This surface notch was aligned with a \hkl{111} glide plane in the sample. To mimic the experimental conditions, the x- and z-directions were taken as open surface boundaries, while the y-direction was set to periodic boundary conditions. This sample geometry is schematically shown in Supplementary Fig.~\ref{figS2}a. 

After equilibration for \SI{50}{\pico\second} at \SI{5}{\kelvin} or \SI{300}{\kelvin} in the NVT ensemble, the sample was subjected to uniaxial strain along y up to a strain of 0.06 using an engineering strain rate of \SI{e8}{\per\second}\cite{Brandl2009}, and a constant time step length of \SI{1}{\femto\second}. Dislocation lines were extracted every \SI{5}{\pico\second} during straining using the dislocation extraction algorithm \textsc{DXA}\cite{Stukowski2010a,Stukowski2012}.

\subsection{Shear of an edge dislocation}

To cut a dislocation into the samples, a half-plane was inserted into an FCC crystal resulting in a misfit edge dislocation following the geometry shown in Supplementary Fig.~\ref{figS2}b\cite{Osetsky2003}. The resulting dislocation had a Burgers vector \(\vec{b}\) along \hkl[1-10] (corresponding to the \(x\)-direction) with a line direction of \hkl[11-2] (corresponding to the \(y\)-direction).

Two different sample dimensions were considered for the isolated dislocation simulations. The smaller sample contained 477600 atoms with periodic boundary conditions along \hkl[1-10] (\(x\)) and \hkl[11-2] (\(y\)) directions and an open boundary along the \hkl[111] (\(z\)) direction. This sample was built from \(200 \times 20 \times 20\) unit cells in each of those directions, respectively. The larger sample had the same crystallographic orientations and boundary conditions, but the total size was increased to \(400 \times 200 \times 20\) unit cells leading to \(\sim \num{11.5e6}\) atoms. In both cases, the lattice sites were filled with the desired atomic species to create a random solid solution.

All samples were annealed for \SI{50}{\pico\second} at \SI{2}{\kelvin} using a \SI{1}{\femto\second} time step. During this annealing and the subsequent shear simulations the surface layers were thermostatted using a Langevin thermostat\cite{Schneider1978}, while the atoms in the center of the sample were integrated in the NVE ensemble.

After equilibration, a ramping shear force was applied to each atom in the outer surface layer. This force pointed in the x-direction (cf. Supplementary Fig.~\ref{figS2}b) and was ramped from 0 to \SI{0.208}{\electronvolt\per\angstrom} over \SI{200}{\pico\second}. From these shear ramp simulations, the critical force to initiate dislocation glide \Fcrit{} was determined.

Once this critical force was established, a second constant force simulation was started. Here, the applied shear force was first ramped from 0 to \Fcrit{} with a rate of \SI{e-6}{\electronvolt\per\angstrom\per\pico\second}.Afterward, it was held for \SI{100}{\pico\second}. 

The dislocation position was extracted using \textsc{DXA}\cite{Stukowski2010a,Stukowski2012} implemented in \textsc{OVITO}\cite{Stukowski2010}.

\subsection{GSF surface calculation}

The global and atomic GSF curves were calculated for the two differently sized samples presented in the previous section before dislocation insertion (Supplementary Fig.~\ref{figS2}b).

For a GSF calculation, the sample is split into two crystallites. One is above the \hkl<111> GSF plane, while the other one is located below this plane. Applying the conventional simulation methodology prevents the atomic relaxation during rigid displacement of the two crystallites within the GSF plane\cite{Vitek1968}, however, the relaxation of the atoms normal to the GSF remains unconstrained. One problem that arose, applying this methodology, was that the atoms in the disordered alloy relaxed into different intermediate states. While this effect averages out in the global GSF calculation, the resulting atomic energy landscapes are not smooth but include discontinuities when an atom relaxes into such a metastable state. To avoid these effects, we further constrained the out-of-plane atomic relaxation during the calculations. During displacement, the forces acting on the atoms along the \hkl[1-10] (\(x\)) and \hkl[11-2] (\(y\)) directions\cite{Vitek1968,Zimmerman2000} were set to 0 while the forces normal to the GSF plane were averaged for all atoms in each crystallite. This means that the pristine FCC lattice was preserved within each crystallite. The crystallites, however, were still able to change their separation distance to reduce the system's energy. Supplementary Fig.~\ref{figS9} shows the difference between this rigid relaxation method compared to the conventional atomic relaxation scheme (relaxed).

Even though this constrained relaxation approach does not resolve the effects of the intrinsic lattice distortions found in HEAs directly, it does show their effect implicitly. A comparison of Fig.~\ref{fig3}a and Supplementary Fig.~\ref{figS8} reveals that the local energy minima in the GSF landscapes do not have to be in the same positions for the global GSF landscape and the ones resolved for each atom, even though both are extracted from the same simulation. The spatial variations in the equilibrium positions for each atom are caused by local fluctuations in chemical environments. In the subsequent calculation of the energy gradients these minima are connected, which implicitly accounts for the lattice distortions.

To determine the GSF landscape, a \(3.6 \times 2.1\)\,\si{\nano\meter\squared} area was sampled with a grid spacing of \SI{0.1}{\angstrom}. At each grid point, the potential energy was minimized using the latest implementation of the \textsc{FIRE} algorithm\cite{Guenole2020} to an energy tolerance of \SI{e-12}{\electronvolt}. The resulting two-dimensional energy landscape for each atom was interpolated using cubic splines and the three local minima were found using the BFGS algorithm (both implemented in \textsc{SCIPY})\cite{Virtanen2020}. The maximum energy gradient was calculated along a linear path connecting these three minima. Fig.~\ref{fig3} shows an example of the resulting atomic energy landscape and calculated energy gradient.

\section{Results}
\subsection{Electron microscopy characterization of dislocation glide and pinning}

\begin{figure*}[tbp]
\includegraphics[width=\textwidth]{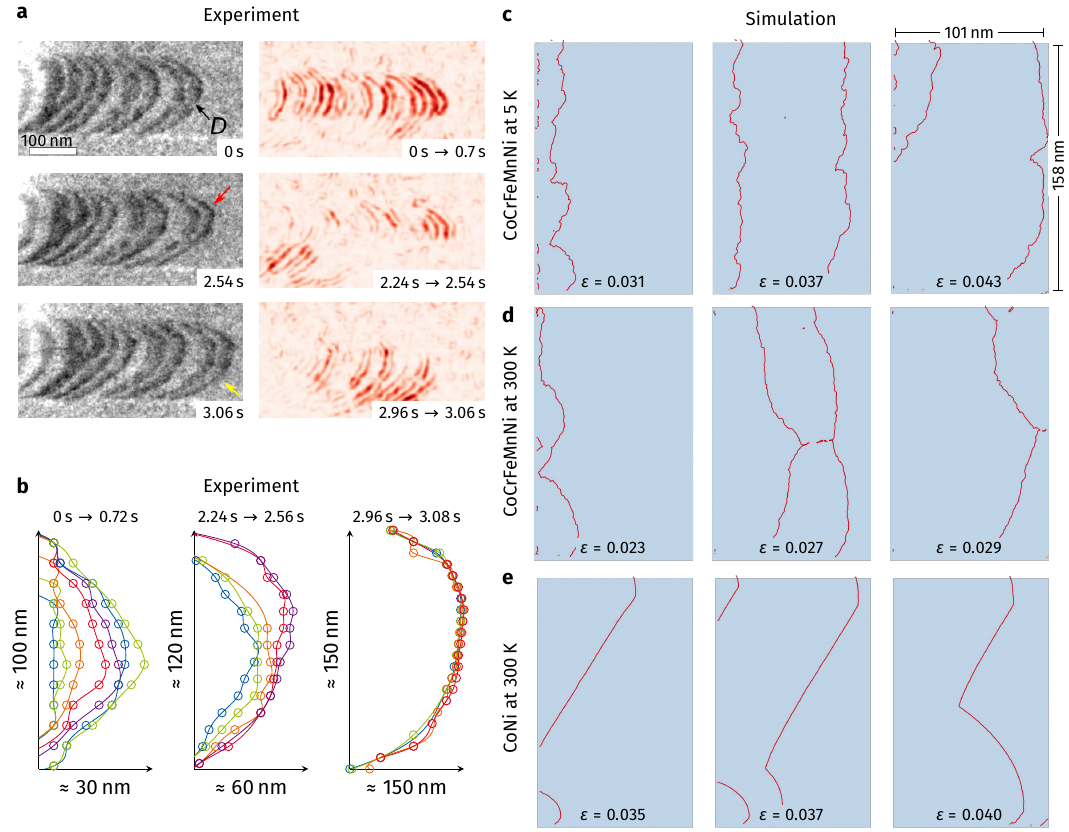}
\caption{%
      Comparison of dislocation lines in the Cantor alloy during in situ TEM straining and atomistic simulations of comparable size.
      \textbf{a} Series of TEM snapshots showing a sequence of gliding Shockley partial dislocations on a \hkl{111}\hkl<110> slip system activated under tensile load. The TEM bright-field image was obtained at two beam diffraction conditions with strong excitation of g022 beam. The SF bound by a pair of leading and trailing partial dislocations is visible with dark contrast. Difference images in the bottom row were prepared by using two TEM frames captured at denoted time intervals. Only a portion of the dislocation lines, marked by the red and yellow arrows, move within the time span.
      \textbf{b} Dislocation line of the leading partial dislocation (marked by ``D'' in \textbf{a}) extracted from consecutive TEM images. The dislocation lines show jerky glide motion; the dislocations exhibit a meandering shape as some segments get pinned during glide. 
      \textbf{c-e} Series of atomistic simulation results showing the strong pinning of dislocations (red) during glide in the equimolar Cantor alloy at cryogenic (\SI{5}{\kelvin}) and ambient (\SI{300}{\kelvin}) temperatures and the weak pinning of dislocations in CoNi alloy at \SI{300}{\kelvin}. Simulation cells with the length scale (\(101 \times 152 \times 75\)\,\si{\nano\meter\cubed}) comparable to the experimental conditions are used. Each sample contains a notch aligned with the shaded \hkl{111} glide plane from which dislocations nucleate during uniaxial straining. For each sample, dislocations are traced at three strain (\(\varepsilon\)) levels. Dislocations in the Cantor alloy show a similar meandering shape and strong localized pinning leading to wavy dislocations as in the experiments whereas dislocations in CoNi alloy show weak pinning, where the dislocation lines remain almost ideally straight.
      }
    \label{fig1}
\end{figure*}

We firstly analyze the structure and glide motion of dislocations in CoCrFeMnNi HEA by in- situ TEM tensile straining (Fig.~\ref{fig1} and Supplementary Movie~1). Uniaxial tensile load is applied to the Cu grid which transfers stress to the TEM lamella prepared by focused ion beam (FIB) lift-out. The loading direction and the viewing direction are \hkl[113] and \hkl[21-1], respectively (Supplementary Fig.~\ref{figS1}) and three \hkl{111} slip planes are inclined with respect to the viewing direction. Due to the finite thickness of the TEM thin foil, which is around \SI{100}{\nano\meter}, dislocation glide on the slip planes is constrained by the top and bottom surfaces.

Fig.~\ref{fig1}a presents the representative TEM snapshots showing the glide motion of an array of Shockley partial dislocations which are separated by SFs appearing as dark bands. The dislocations bow out under the local resolved shear stress toward their glide direction. All images are processed to enhance the dislocation contrast (see section “Methods”). Tracing the evolution of the dislocation line shape during glide reveals that dislocations move through a kink-pair-like mechanism\cite{Caillard2003}. For example, only the upper part of the dislocation marked by a red arrow moves forward, then with some delay, the bottom part (yellow arrow) follows. The difference images in the right panel of Fig.~\ref{fig1}a , highlighting the relative changes between two frames, show that only certain segments of dislocations move within a given time frame while others remain pinned.

This jerky motion can be seen more clearly by tracing the dislocation position over time (Fig.~\ref{fig1}b). The positions are extracted using Gaussian and ridge detection image filters. As described above, a short segment of the dislocation line propagates first by forming a small hump, while the rest hardly moves. Most dislocations exhibit this type of localized pinning-unpinning along their lines constantly during glide, which could be the reason for the high shear stress required for dislocation glide in HEAs\cite{Lee2020}. The effects of experimental artifacts, such as FIB damages or surface oxide, are expected to be marginal since these are present at the outer surface of the TEM sample with a small thickness of \(5-10\)\,\si{\nano\meter}\cite{Kiener2007,Mayer2007} while the total thin foil thickness is around \SI{100}{\nano\meter}. Therefore, the jerky motion in the middle of a dislocation, which is placed near the center of the TEM foil should not be affected by these surface artifacts.

In order to investigate whether the inherently meandering dislocation line shape seen in TEM can be reproduced via atomistic computer simulations, samples of comparable size are built. Each sample contains about 100 million atoms and features an inclined glide plane measuring \(100 \times 160\)\,\si{\nano\meter} (shown in Supplementary Figs.~\ref{figS2}a,c-d). To facilitate dislocation nucleation under tensile load a surface notch, aligned with this glide plane, is cut into the sample before loading the sample. Figs.~\ref{fig1}c-e show the first Shockley partial dislocations emitted and gliding away from this notch under tensile load. At low temperatures (\SI{5}{\kelvin}), strong pinning of the dislocation line in the HEA can be seen (Fig~\ref{fig1}c). These manifest as “V”-shaped dislocation line arrangements, where the line is pinned in one spot, while neighboring dislocation segments advance. These pinning points occur frequently, which results in a rough dislocation line. Thermal activation at ambient temperature renders dislocations apparently straighter (\SI{300}{\kelvin} Fig.~\ref{fig1}d). The thermal energy lifts the dislocation over some of its weaker pinning sites, thus only the strong pinning points, which have a much lower density, persist active. For comparison, dislocations in the binary CoNi alloy were studied in an identical geometry (see Fig.~\ref{fig1}e). There are almost no dislocation pinning points in this material, leading to straight dislocation segments during glide. The cusp in the middle of the dislocation stems from the sequence of initial nucleation events at the notch (cf.\ Supplementary Fig.~\ref{figS2}d,e). Thus, the effect of strong dislocation pinning in HEAs and weaker dislocation pinning in binary alloys can be resolved. Moreover, the molecular dynamics (MD) simulations provide clear evidence for intrinsically wavy dislocations in the absence of SRO as observed, experimentally.

\begin{figure*}[tbp]
  \includegraphics[width=\textwidth]{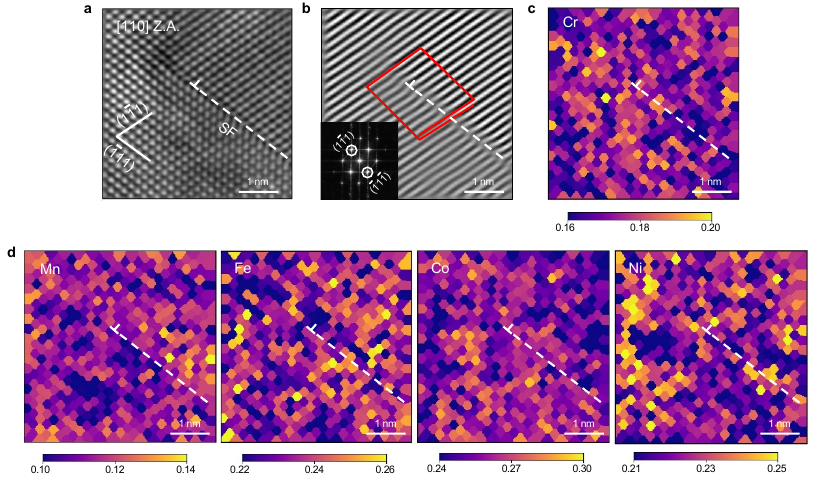}
  \caption{%
      STEM EDS composition maps around a Shockley partial dislocation core.
      \textbf{a} STEM HAADF image with \hkl[110] zone axis showing a Shockley partial dislocation. The dislocation core is indicated by symbol \(\bot{}\) and its stacking fault (SF) by a dashed line. 
      \textbf{b} Bragg filtered image of selected \hkl(1-11) and \hkl(-11-1)reflections. Burgers circuit—showing the Burgers vector of \(a_0/6\hkl[211]\).
      \textbf{c, d} EDS composition map of Cr, Mn, Fe, Co, and Ni around the dislocation core showing a homogeneous distribution with random fluctuation but without enrichment or clustering of any specific element. Note the differences in color scale for each element. A detailed analysis of the dislocation is given in Supplementary Fig.~\ref{figS7}.
  }
  \label{fig2}
\end{figure*}

A comparison of experiment and simulation in Fig~\ref{fig1} is only valid if both samples are in a similar random solid solution state. While it is clear from the sample preparation that there is no elemental clustering in the virtual sample, this question has not been answered definitively for the TEM sample. Lee et al.\cite{Lee2020} previously performed atom probe tomography measurements on the samples shown in Fig~\ref{fig1}a and found no elemental ordering or clustering on the length scale of \(\sim\SI{1}{\nano\meter}\). For further detailed analysis, we carried out atomic-resolution STEM EDS chemical mapping and quantitative analysis combined with EDS simulations (Supplementary Figs.~\ref{figS3}-\ref{figS6}) to determine the concentration of each atomic column around a Shockley partial dislocation. The high-angle annular dark field (HAADF) STEM image of the Shockley partial and its Bragg filtered analysis are given in Fig~\ref{fig2}a,b, confirming the \(a_0/6\hkl[211]\) Burgers vector of the selected dislocation core, where \(a_0\) is the lattice constant. With this quantitative evaluation of composition at each atomic column combined with EDS simulation, we could conclude the chemical homogeneity around the dislocation core without apparent clustering or segregation of specific element(s). There is fluctuation in per-column concentration though, which is, however, in line with the pristine bulk and negligibly small (Supplementary Fig.~\ref{figS4}). This result is also consistent with the recent high-resolution EDS experiments by Heczko et al.\cite{Heczko2021}, who report no elemental segregation to the strain field of Shockley partial dislocations in heat-treated Cantor alloy.

\subsection{A local descriptor for dislocation pinning in HEAs}

Fig.~\ref{fig1} shows an inherent waviness of dislocation lines during glide in both experiment and simulation, indicating the existence of dislocation pinning sites. While we know that the simulated sample is in a perfectly random state without elemental clustering or short-range order, the experiments also show that there is no clustering or segregation around the dislocation core of Shockley partial dislocation.

Therefore, some statistically occurring atomic arrangements must be present, even in the random alloy, which pin the dislocation lines and locally hinder their glide. In the following, we investigate a single (dissociated) edge dislocation in the equimolar CoCrFeMnNi HEA to determine the nature of these nanoscale dislocation pinning points.

The established Peierls model of dislocation glide states that the stress required to move a dislocation is proportional to the maximum gradient along the energy path connecting two adjacent stable dislocation positions. For pure metals, this energy pathway (Peierls energy) may be approximated by a sinusoid and the so-called Peierls stress can be calculated analytically as a derivative of this energy. In dilute solid solutions, the local energy landscape is modified in the vicinity of each solute and therefore the local stresses required for dislocation glide changes\cite{Hirth1982,Bulatov2006}.

As the exact form of the Peierls energy landscape is usually unknown, the generalized stacking fault (GSF) energy surface\cite{Lu2000,Vitek1968} is often taken as a stand-in. The GSF energy surface maps the energy required to rigidly displace two densely packed crystal planes against each other. Supplementary Fig.~\ref{figS8} shows the GSF energy surface for a HEA sample. Here, the displacement and energy profiles corresponding to leading and trailing partial are indicated. As the GSF energy profile is obtained by averaging over the whole lattice plane, it is symmetric and smooth, showing no effect of the local chemical inhomogeneity. However, given the strong variance in local chemical arrangements, it is expected that the GSF energy varies on a local scale in HEAs.

It is widely accepted that the stable SF energy in HEAs, which marks a single point on the GSF curve strongly depends on the chemical environment\cite{Ding2018,Ikeda2018}. Therefore, one would expect this dependence also for other points on the GSF landscape. From the locally fluctuating GSF landscape, one can directly conclude that the Peierls stress and thereby resistance against dislocation glide in HEAs becomes a spatially varying quantity. In the following, we will calculate this for each atom in the glide plane.

While the calculation of a local GSF energy landscape seems counterintuitive at first glance, it becomes a rather well-defined problem in the framework of classical atomistic simulations. LAMMPS\cite{Thompson2022}, by relying on interatomic potentials, is not only able to calculate the total energy for an arrangement of atoms but also assign a specific energy contribution to each atom. If we now perform a GSF energy calculation for the HEA system, we can store these per-atom energies and therefore obtain not only the global GSF energy landscape but also one resolved for each atom. While the general idea of this approach is similar to the stochastic Peierls-Nabarro model proposed in Refs.~\cite{Zhang2019,Jiang2020}, we do not attempt to find a closed form mathematical description of the local Peierls stress but instead measure the atomic Peierls force directly and confirm their effect on the dislocation pinning in a given alloy.

\begin{figure*}[tbp]
  \includegraphics[width=\textwidth]{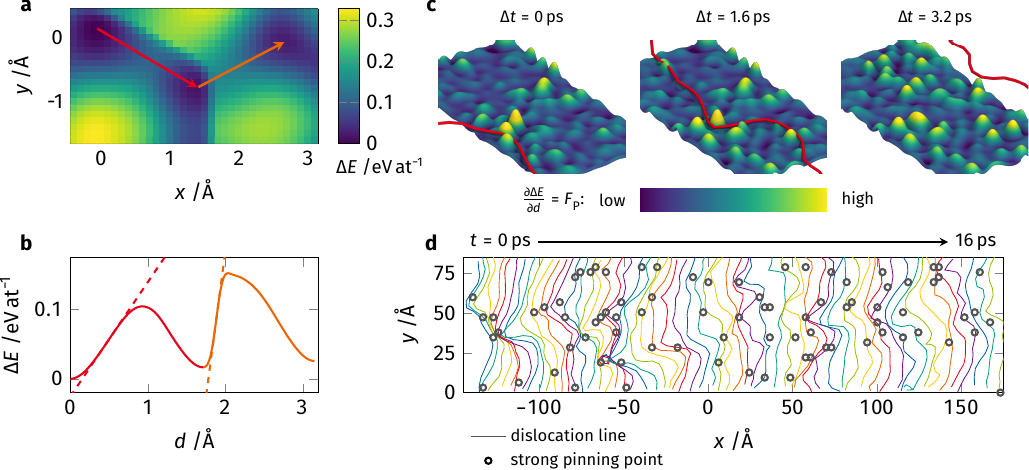}
  \caption{%
      Atomistic Peierls barrier as a descriptor for dislocation pinning --- dislocation in CoCrFeMnNi moves through this pinning point landscape.
      \textbf{a, b} GSF energy projected on a single atom in the glide plane. Red and orange arrows indicate the respective displacements connecting the local minima. The local Peierls barrier \Fp{} (in units of force) is defined as the maximum slope of this atomic energy curve. For reference, the GSF landscape averaged over the whole sample is given in Supplementary Fig.~\ref{figS8}.
      \textbf{c} Representation of the local Peierls force in the glide plane (color corresponds to the local pinning force) with the dislocation (red) superimposed during glide. The per-atom \Fp{} values are averaged on a \SI{3}{\angstrom} by \SI{3}{\angstrom} grid and smoothed using Gaussians for visual representation. It can be seen that the dislocation is pinned on the strongest obstacle, bends around it, and finally detaches. Note how the displacements caused by the passing dislocation change the pinning point landscape for subsequent dislocations. The mapped region spans \SI{45}{\angstrom} by \SI{90}{\angstrom}. 
      \textbf{d} Superposition of different snapshots of the leading partial dislocation during glide. As snapshots are taken at constant time intervals, dislocation lines being closely together correspond to long dwell times, i.e., stronger localized pinning. The position of the strongest dislocation pinning points is marked by gray circles. Good agreement between the two can be seen.
  }
  \label{fig3}
\end{figure*}

Fig.~\ref{fig3}a shows the energy landscape for a single Co atom in the GSF plane. Compared to the globally averaged GSF energy landscape (Supplementary Fig.~\ref{figS8}) one can already see that the projected atomic energies are not symmetric anymore, as the local chemical environment changes after the initial displacement (red). Therefore, displacing this Co atom along the leading (red) and trailing (orange) partial dislocation displacement path does not result in the same energy barrier. Fig.~\ref{fig3}b shows the corresponding energy profiles along the two displacement paths. Dashed lines indicate the maximum gradient of the GSF curve, which corresponds to the local Peierls friction force. This value is proportional to the strength of this atom acting as a pinning point for the dislocation line and is denoted as \Fp{}. Note, this pinning point's strength is given in units of force (\si{\electronvolt\per\angstrom}). It cannot be normalized to stress (\si{\electronvolt\per\angstrom\cubed}) as the area per-atom is not well defined in HEAs.

The next question is whether this approach actually works in describing the dislocation pinning on the atomic scale. To answer this, we insert a single edge dislocation into this pre-calculated GSF plane and apply a constant shear load to it. Fig.~\ref{fig3}c shows the leading partial dislocation of the dissociated edge dislocation gliding from left to right across the calculated pinning point landscape (the atomic pinning point strength \Fp{} is averaged on a \(3 \times 3\)\,\si{\nano\meter\squared} grid and smeared out for better visibility). Movie~S2 shows this sequence in motion. Here, the dislocation is pinned on the highest strength pinning point, i.e., the strongest, obstacle and bows out under the applied shear. Once the adjacent dislocation line segments have advanced a sufficient distance, the increase in line tension lifts the dislocation line across this strong obstacle and the dislocation continues its glide. After the dislocation has passed a given area the atoms are shuffled and the dislocation pinning point landscape changes for following dislocations.

Lastly, Fig.~\ref{fig3}d shows the extracted position of the leading partial dislocation during glide over a \SI{16}{\pico\second} time interval. As the dislocation snapshots are taken every \SI{250}{\femto\second}, multiple closely spaced line segments correspond to the positions where the dislocation line is locally pinned while wider line spacing indicates faster dislocation glide. Superimposed are gray circles indicating positions of the highest local pinning point strength predicted from the atomic GSF calculations. Even by eye, a good match between proposed pinning points and regions of strong pinning is discernible. Obviously, there is no simple 1:1 correspondence between dislocation line shape and proposed pinning sites. This is the case as the “history” of the dislocation changes its current behavior. If the dislocation line was initially pinned and is therefore strongly bowed out, there is an additional force arising from the line tension to straighten out the dislocation line and shorten its overall length which can assist the crossing of pinning points along its way.

Given that our simulations have found an atomic scale descriptor for dislocation pinning points that is in line with the established theory of the Peierls model, we test this descriptor for a wide range of HEAs and search if an increased number of high \Fp{} lattice sites occurs.

\begin{figure*}[tbp]
  \includegraphics[width=\textwidth]{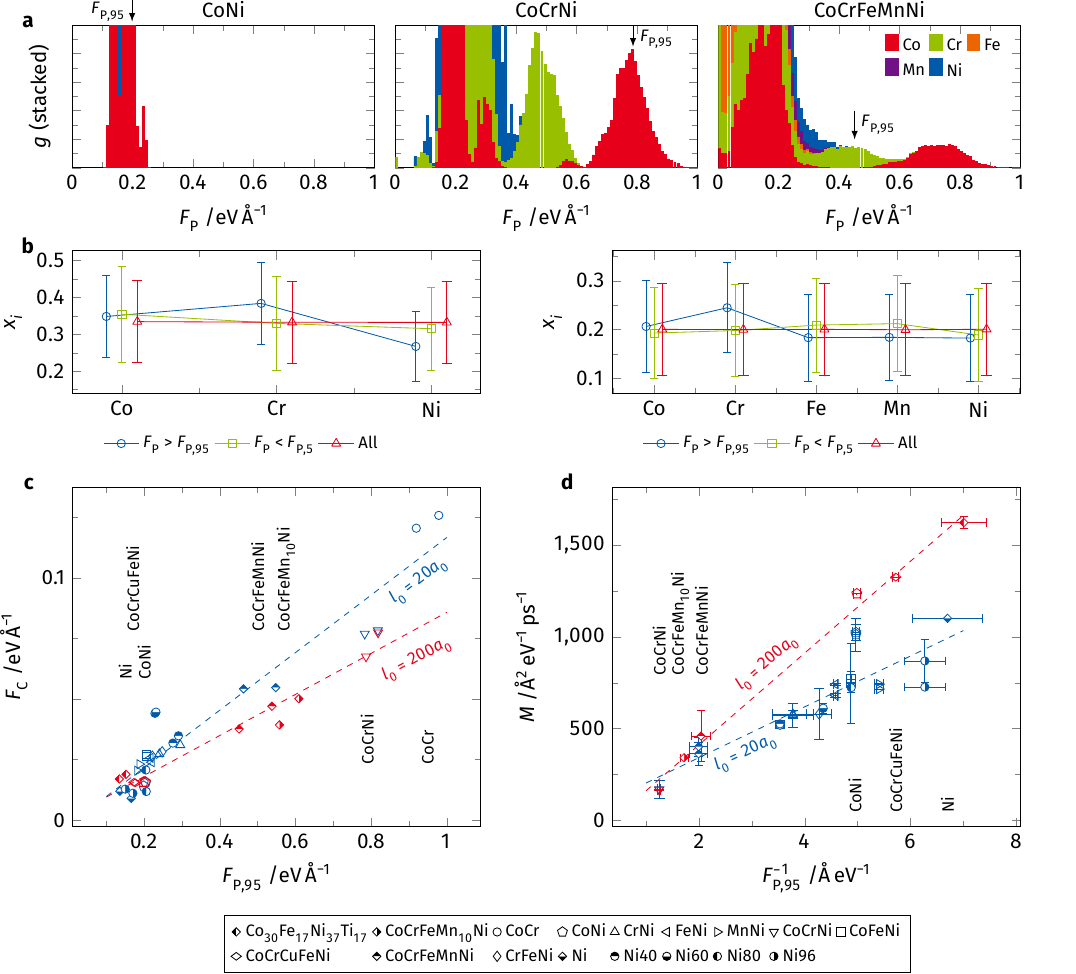}
  \caption{%
      Distribution of the atomic pinning point strength in different alloys and their relation to the critical force for dislocation glide and dislocation mobility.
      \textbf{a} Comparison of the element resolved pinning point strength \Fp{} in the equimolar CoNi, CoCr Ni, and CoCrFeMnNi. The interaction of Co and Cr leads to the appearance of a high density of strong pinning points.
      \textbf{b} Mean composition xi in the first and second nearest neighbor shell around strong (\(\Fp{} > \FpP{}\)) and weak (\(\Fp{} < F_{\text{P},5}\)) pinning points in the CoCrNi and CoCrFeMnNi alloys. Note, that the species of the central atom (exclusively Co for the strong pinning points) is excluded in this concentration measure. Error bars correspond to the standard deviation in the concentration found around all atoms adjacent to the glide plane. 
      \textbf{c} Correlation of the critical force \Fcrit{} required to initiate dislocation glide in the different samples (\(l_0\) denotes the initial dislocation line length in units of the lattice constant \(a_0\)) and the high strength pinning point density \FpP{}.
      \textbf{d} Mobility \(M\) of the dislocation under constant applied force \Fcrit{} plotted against the inverse of the high strength pinning point density \(F^{-1}_{\text{P},95}\). Error bars highlight the differences in \FpP for encountered by leading and trailing partial dislocation as well as to the standard deviation of mobilities measured during the simulation run. Only selected samples are highlighted for clarity.
  }
  \label{fig4}
\end{figure*}

Following the steps outlined in the previous section, we prepare a wide range of samples of different composition ranging from binary and ternary alloys to different HEAs. First, the atomic GSF landscape is calculated for all atoms in a given \hkl{111} plane. Fig.~\ref{fig4}a shows the calculated \Fp{} distribution \(g(\Fp{})\) for CoNi, CoCrNi, and CoCrFeMnNi. This comparison reveals that in CoNi almost all pinning points have a resistive force below \SI{0.2}{\electronvolt\per\angstrom}. The addition of Cr leads to a substantial change in this spectrum. The interaction of Co and Cr causes the emergence of a new peak with \Fp{} values \(\sim\SI{0.8}{\electronvolt\per\angstrom}\) where Co atoms have become substantial obstacles for the dislocation line. Some Cr atoms occupy states of intermediate pinning point strength. Subsequent addition of Fe and Mn, on the other hand, does not alter the observed pinning point spectrum much. The strength of the Co and Cr atoms remains unchanged but as their concentration decreases so does the density of strong pinning sites. Both Fe and Mn are only weakly pinning (\(\Fp{}\approx\SI{0.2}{\electronvolt\per\angstrom}\))elements within this matrix.

The three examples presented in Fig.~\ref{fig4}a suggest the Co-Cr interaction to be responsible for strong dislocation pinning points. To confirm, we explore the mean concentration in the first and second nearest neighbor shell of the strongest (95th percentile \FpP{}) and the weakest (5th percentile \(F_{\text{P},5}\)) pinning points. The resulting concentrations, given in Fig.~\ref{fig4}b, reveal that the concentration of Cr is increased around the strongest pinning points. Be reminded that the strong pinning points are exclusively Co atoms, confirming the interaction of these two species. The weak pinning points, on the other hand, mirror the bulk concentration (labeled “All”) suggesting that there are distinct atomic environments responsible for the strongest dislocation pinning, even in a random alloy. However, the chemical concentrations of the atomic coordination acting as strong pinning sites show a great variance suggesting that there is not a single atomic arrangement responsible for the dislocation pinning.

Next, we investigate the effect of the pinning point spectrum on the critical force required to initiate dislocation glide as well as their mobility. One would expect dislocation pinning to be dominated by the fraction of strongest pinning points holding back the dislocation line. Therefore, we take \FpP{} as a descriptor. For reference, this 95th percentile is marked in Fig.~\ref{fig4}a. The dislocations in each sample are subjected to increasing shear until they start to glide. The critical applied force to initiate glide, \Fcrit{}, is determined as the force required to move the dislocation by \SI{20}{\angstrom}. These simulations are carried out at cryogenic temperatures to reduce thermal noise on the dislocation line shape. Note that this force is not necessarily identical for leading and trailing partial dislocations, especially for CoCrFeMnNi and CoCrNi which have negative SF energies, resulting in an additional driving force acting on the leading partial. Fig.~\ref{fig4}c shows \Fcrit{} over \FpP{} for all samples. Blue symbols indicate results for the smaller samples with an initial dislocation line length \(l_0\) of 20 unit cells, while red symbols indicate results from larger samples with an initial line length of 200 unit cells, i.e., lattice constants \(a_0\). A clear linear relationship between the pinning point density and the critical force can be seen within the small and large sample series confirming the expected correlation, where an increased density of strong pinning points leads to an increased force required to move a dislocation. Note that both sample sets are not on the same line, as the longer dislocation has a higher probability of having a locally weakly pinned segment where glide can start. This segment subsequently drags the other dislocation segments along. The mechanism is comparable to the weakest-link model proposed by Nöhring and Curtin\cite{Nohring2018} for cross-slip in FCC HEA, as suggested by Rizzardi et al.\cite{Rizzardi2018}.

We also calculate the dislocation mobility under a constant applied shear (Fig.~\ref{fig4}d). Here, each sample is subjected to constant shear equivalent to the critical force required to move the leading partial dislocation by \(10-30\)\,\si{\angstrom} (depending on the sample). In some samples, repeated pinning and depinning of the dislocation line were observed. Here, we isolated segments of constant velocity v and calculated the mobility \(M\) as, \(\nu = MF\)\cite{Hirth1982}, where \(F\) is the applied shear force. While the critical force was found to be proportional to \FpP{}, the dislocation mobility should be inversely related to this value—the stronger the pinning points are, the slower and less mobile the dislocation will be. Therefore, we plot the extracted dislocation mobilities against \(F^{-1}_{\text{P},95}\). Again, a linear correlation between the dislocation mobility and the pinning point descriptor can be seen within each set of samples. Similar to the critical force, the weakest-link argument can be taken as an explanation for the differences between the two differently sized samples. Recent work by Sills et al.\cite{Sills2020} showed a similar effect in stainless steels, where very long dislocation line lengths were required to converge the calculated dislocation mobility. However, comparing samples of identical initial dislocation line lengths should be valid.

To confirm that the correlations shown here are no coincidence, we also compare \Fcrit and \(M\) against other descriptors derived from the per-atom GSF curves. These can be found in Supplementary Figs.~\ref{figS12} and \ref{figS13}. Here we show the atomic stable and unstable stacking fault energies to \Fp{}, plotting their distribution following Fig.~\ref{fig4}a and the resulting correlations akin to Figs.~\ref{fig4}c,d. This confirms that \Fp{} is by far the best descriptor for \Fcrit{} and \(M\).

In summary, we show that the dislocation pinning points identified in the Cantor alloy could also be found in different binary and ternary alloys. The critical force required for a dislocation to move correlates with the density of high strength pinning points identified from the atomic scale resolved GSF curves. Similarly, a greater number of strong pinning sites leads to overall reduced dislocation mobility in all investigated samples, independent of the material system or interatomic potential used. Investigation of the chemical environment around each pinning point reveals Co at its center and an increase in Cr in its surrounding, however, no single structural motif responsible for dislocation pinning has been confirmed. Given that the interaction of Co and Cr atoms needs to be harnessed, direct adjacency seems to be preferable.

\section{Discussion}
\subsection{Impact of the dislocation pinning points}

We propose a descriptor for dislocation pinning sites in concentrated alloys. The descriptor is physically motivated and based on the atomic Peierls friction, which can be calculated as the maximum gradient of the per-atom GSF energy. We calculate the descriptor for a given pristine lattice plane, which is subsequently traversed by a dislocation allowing for direct spatial correlation of the established descriptor and observed dislocation pinning sites. Concentration analysis of the strongest dislocation pinning points reveals that they stem from the unique, nonlinear, interaction of Co and Cr atoms.

\begin{figure*}[tbp]
  \includegraphics[width=\textwidth]{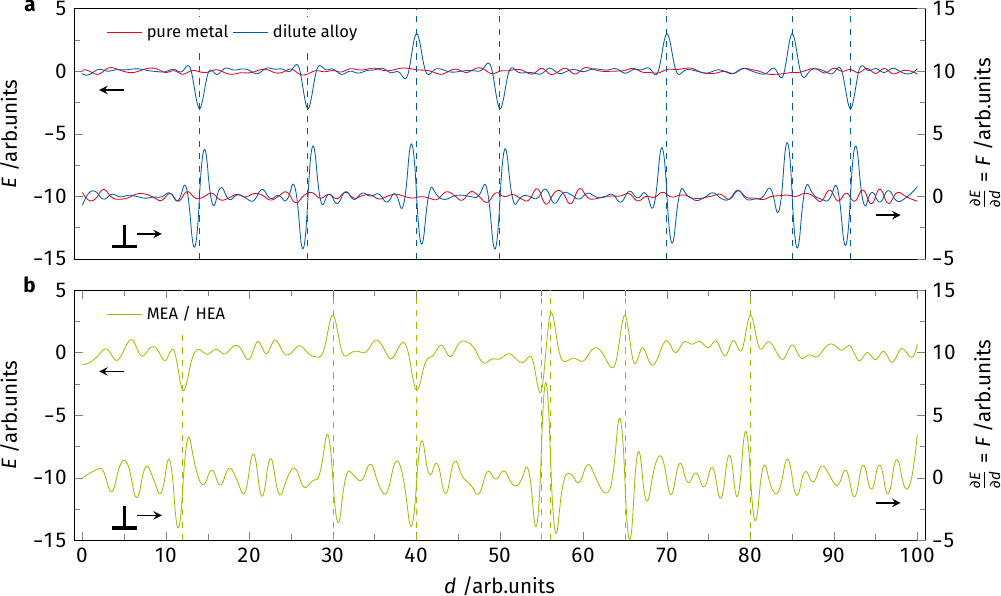}
  \caption{%
      Schematic representation of the energy and friction force landscape encountered by a dislocation.
      \textbf{a} Energy change \(E\) as a dislocation glides distance \(d\) in a pure metal or a dilute alloy. In pure metal, the energy remains almost constant during dislocation glide as all chemical environments are identical. The dilute solid solution shows a similar energy profile as the pure metal away from any solute atoms. Close to the solutes (marked by dashed lines) the dislocation can be pinned either by local energy minima from which it needs to escape or by higher energy barriers. The required force \(F\) (gradient of the energy) is shown on the right \(y\)-axis. An increased glide resistance can be seen around the energy walls and valleys.
      \textbf{b} shows \(E\) and \(F\) for concentrated HEAs. Here, the variance in \(E\) and \(F\) is higher in the matrix. However, a combination of adjacent local minima and maxima (\(d=55\)) leads to a much higher variation in local pinning forces and higher maxima.
  }
  \label{fig5}
\end{figure*}

The energy and force landscape felt by a dislocation is schematically shown in Fig.~\ref{fig5}. In a pure metal, the energy change \(E\) encountered by a dislocation during glide is negligibly small. Therefore, the gradient of this energy and hence the force \(F\) required to move the dislocation is almost zero as well\cite{Gilman2007}. If a low concentration of solutes is added to the metal, forming a dilute solid solution, these solutes can either form local energy minima that trap the dislocation or high energy barriers blocking dislocation advancement. Both effects lead to steep energy gradients and therefore require high forces to be passed (Fig.~\ref{fig5}a). In a concentrated HEA (Fig.~\ref{fig5}b), on the other hand, the energy variance in the matrix is already higher hindering dislocation glide. However, as shown in Figs.~\ref{fig3} and \ref{fig4}, there are some environments that pin dislocations exceptionally strongly. These could stem from atomic arrangements where an energy minimum is located next to a barrier (as shown around \(d=55\) in Fig.~\ref{fig5}b), leading to a steeper energy gradient. While these arrangements are theoretically also possible in dilute alloys, their probability is much lower given the lower density of secondary atoms. To confirm this schematic image, we also prepared linear profiles of the \Fp{} landscape in Co, Co\(_{99}\)Cr\(_{1}\), and CoCrFeMnNi samples. These are shown in Supplementary Fig.~\ref{figS11} and confirm that there are no strong pinning points in the pure metal, some discrete strong pinning sites in the dilute CoCr solid solution and an abundance of pinning spots in the full HEA. The strong pinning spots in the CoCrFeMnNi alloy do not only show an increase in frequency but also higher peak \Fp{} values.

The identification of dislocation pinning points allows us to sample their density for different binary and ternary alloys and compare those to the CoCrFeMnNi HEA. As shown in Fig.~\ref{fig4}a, neither Co nor Cr alone leads to a substantial shift in the \Fp{} distribution when added to Ni. Only the addition of both gives rise to a pronounced increase in pinning point density and strength. Further addition of Fe and Mn leads to no relevant changes in the \Fp{} spectrum and therefore only decreases the density of strong pinning points. The intuitive trends expected from the densities of strong pinning points, i.e., increasing resistance against dislocation glide and reduced mobility as the fraction of strong pinning points in a material increases are confirmed by our single dislocation shear simulations. Even though this is by no means a solid solution strengthening model, one might naively assume that an increased resistance against the onset of dislocation glide corresponds to an increase in the strength of a material. Under this assumption, our findings would be in line with experimental observations, where CoCrNi consistently shows higher hardness\cite{Wu2014} and tensile yield strength\cite{Wu2014a,Slone2019} compared to other alloys from the same family, including the \num{5} component CoCrFeMnNi Cantor alloy. Additional simulations show that Ni is not even a necessary constituent, as a FCC random solid solution of CoCr has an even higher dislocation pinning than CoCrNi. This is in agreement with established literature, where FCC Co\(_{63}\)Cr\(_{29}\) alloys are commonly used in dental prosthetics due to their high yield strength and work hardening\cite{AlJabbari2014}.

Li et al.\cite{Li2019d} reported the strengthening in regions of increased Co-Cr bonding by calculating the activation energy for dislocation motion in samples with different degrees of Co-Cr local ordering. Their calculation method requires the use of dedicated smaller samples, which means that they cannot calculate the energy barrier for long-range dislocation glide. Our approach, on the other hand, allows the determination of the atomic pinning point strength in the glide plane of the dislocation. Therefore, a direct spatial correlation between the atomic descriptor and pinned dislocation line can be made. Antillon et al.\cite{Antillon2019} also proposed a method of calculating the local energy barrier against dislocation glide. Their method, however, relies on energy measurements as the dislocation passes through the material in a MD simulation. In contrast, the proposed descriptor can be calculated from the pristine material, i.e., without an embedded dislocation, under static conditions. Therefore, this descriptor has the possibility of being used for future materials design, as it does not require costly MD simulations of many hundred-thousands of atoms to identify potential chemical compositions of strong dislocation pinning.

\subsection{Transferability of our results}

The simulation results outlined in this manuscript were all obtained at cryogenic temperatures. This suppresses the thermally assisted passing of dislocation pinning points. Within the framework of transition-state theory, the energy barrier (not the pinning forces considered in Figs.~\ref{fig3} and \ref{fig4}) associated with a pinning point becomes important to determine whether a dislocation is pinned on a given pinning site. Therefore, the pinning points found here will only pin a dislocation, if the applied force is below the pinning point strength and the associated energy barrier is greater than the thermal energy. Assuming that the energy landscape is unchanged by both temperature and the presence of the dislocation, we can take the per-atom unstable stacking fault energy as a surrogate for the local energy barrier. Considering our data, we can see that the unstable stacking fault energy shows broad distributions in these alloys: CoNi \SI{73.2+-16.7}{\milli\electronvolt}, CoCrNi \SI{100.4+-32.9}{\milli\electronvolt}, and CoCrFeMnNi \SI{74.7+-47.7}{\milli\electronvolt}, with overall relatively low energy barriers suggesting that thermal activation will become important at intermediate temperatures.

We propose that the dissociated edge dislocation studied here is representative of FCC materials. In general, the calculated \Fp{} values are intrinsically insensitive to the dislocation character, as they are calculated based on the pristine lattice. This also extends to the actual pinning point-dislocation interaction. Our descriptor only accounts for the rigid displacement of the atoms during the passage of the dislocation. This means that even though the shear direction (in relation to the line direction) is different for screw and edge dislocations, the shear process is comparable. Hence, there is no obvious dependence of our descriptor on the dislocation character. However, the displacement paths that need to be calculated to obtain \Fp{} depends on the shear direction, i.e., the Burgers vector. This is an obvious difference from established solid solution strengthening models, e.g., Ref.\cite{Varvenne2016a}, where the elastic interaction between solute and dislocation depends on their stress fields and therefore on the dislocation character. Based on these arguments, we also expect the coupling of Co and Cr, shown in Fig.~\ref{fig4}b, to be independent of the dislocation character.

Dislocations in body-centered cubic (BCC) metals glide differently than dislocations in FCC metals. Usually, the Peierls barrier in BCC metals is much higher requiring a kink-pair mechanism to advance the dislocation\cite{Hirth1982}. Based on this mechanism, the displacement of the atoms is most likely different from the rigid shear displacements covered by our \Fp{} descriptor. Therefore, it is most likely not directly applicable. However, if one finds the rigid shear ``path'' or trajectories of the atoms involved in such a process in the BCC matrix, the resulting force barriers could be calculated in a similar way. The glide of screw dislocations through a BCC HEA matrix has recently been studied in detail by Wang et al.\cite{Wang2022}, who use nudged elastic band calculations to understand the energy landscape of the gliding dislocation through the BCC matrix.

Our study reveals the atomic origin of dislocation pinning points in multi-principal element alloys thereby answering a long-standing question in the materials science and engineering community. We find that no distinct SRO is required for strong dislocation pinning in HEAs as local pinning of the dislocation line is facilitated by a locally increased slope of the GSF curve. This scenario agrees with the established Peierls model but reveals that HEA and specific alloys are susceptible to exceptional Peierls peak heights. In the alloy samples investigated, these changes are highly non-linear and the interaction of Co and Cr leads to the highest density of strong dislocation pinning points. This understanding of the local Peierls stress landscape and the resulting pinning forces acting on dislocations gives access to a different parameter that can be tuned in the design of alloys with increased resistance against the onset of dislocation glide and overall reduced dislocation mobility.

\section*{Acknowledgements}

The authors would like to acknowledge financial support by the Deutsche Forschungs Gemeinschaft (DFG) under Grant Nos.\ AL 578/25-2, and DE 796/13-1 as part of SPP 2006. S.L.\ is funded by Alexander von Humboldt Foundation. S.H.O.\ was supported by the KENTECH Research Grant (KRG2022-01-019), National Research Foundation of Korea (NRF) grant funded by the Korean government (MSIT) (NRF-2020R1A2C2101735), Creative Materials Discovery Program (NRF-2015M3D1A1070672, NRF-2019M3D1A1078296), Bioinspired Innovation Technology Development Project (NRF-2018M3C1B7021994). Calculations for this research were conducted on the Lichtenberg high-performance computer of the TU Darmstadt. The authors gratefully acknowledge the Gauss Centre for Supercomputing e.V. (www.gauss-centre.eu) for funding this project by providing computing time on the GCS Supercomputer SuperMUC-NG at Leibniz Supercomputing Centre (www.lrz.de). Author contributions S.H.O., G.D., and K.A.\ conceived the research. D.U.\ and A.S.\ carried out and analyzed the computational simulations. S.L.\ carried out in situ TEM experiment and the analysis of TEM data. H.J.\ and Y.X.\ conducted STEM EDS experiments and analysis. The first draft of the paper was written by D.U.\ and S.L., and all authors contributed with comments and edits.

\bibliographystyle{unsrtnat}
\bibliography{reference}

\cleardoublepage
\renewcommand{\thefigure}{S\arabic{figure}}
\renewcommand{\theHfigure}{Supplement.\thefigure}
\setcounter{figure}{0}

\section*{Supplementary Figures}

\begin{figure*}[htbp]
  \includegraphics[width=\textwidth]{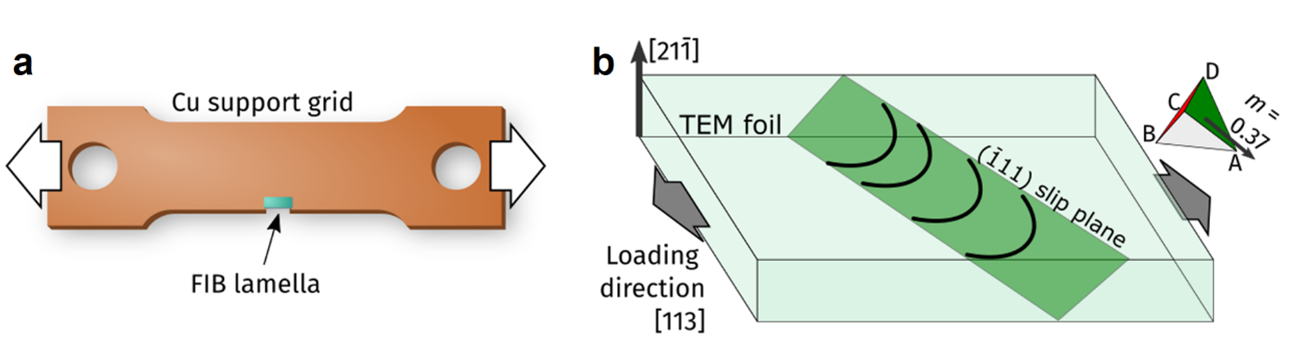}
  \caption{%
      Geometry of in-situ TEM samples.
      \textbf{a} Cu supporting grid for in-situ straining holder. A small notch is cut by FIB milling, and the FIB lamella was positioned on top of it. 
      \textbf{b} A schematic drawing showing crystallographic orientation of the sample. Uniaxial load which was directly applied to the Cu inset, also strained the TEM foil. Dislocations in Fig~\ref{fig1}a glided on the \hkl(-111) plane which was inclined with respect to the electron beam direction.
  }
  \label{figS1}
\end{figure*}

\begin{figure*}[htbp]
  \includegraphics[width=\textwidth]{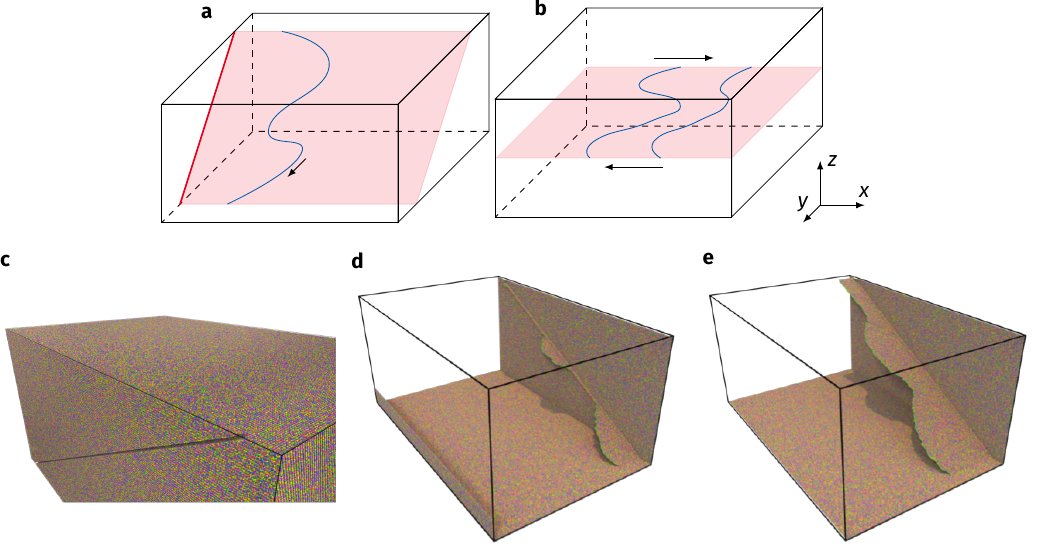}
  \caption{%
  Schematic drawing of the samples used the atomistic computer simulations.
  \textbf{a} Simulated TEM lamella used to investigate the dislocation nucleation from a surface notch (red line). Shockley partial dislocations (blue line) nucleate under the applied tensile strain (arrow) and move on the \hkl{111} glide plane (red). The sample has periodic boundary conditions along the y-direction with open boundaries along \(x\)- and \(z\)-directions. 
  \textbf{b} Sample setup used to determine the critical force \Fcrit{} for dislocation glide and the dislocation mobility \(M\). Shockley partial dislocations (blue) adding up to a perfect edge dislocation are inserted on a \hkl{111} glide plane (red). They start to glide under the applied shear force on the top and bottom surface layers (arrows). The sample has open boundaries along \(z\)-direction, while the two other directions have periodic boundary conditions. Further details are given in the methodology section. 
  \textbf{c} Close-up of the notch cut into the large CoCrFeMnNi sample after equilibration at \SI{5}{\kelvin} for \SI{50}{\pico\second} prior to straining. 
  \textbf{d-e} Dislocation nucleating from the notch under applied strain. The dislocation line determined from Dislocation analysis (DXA), implemented in OVITO is shown in green, non-FCC atoms are hidden for better visibility of the leading partial dislocation and resulting stacking fault (SF). Atoms on the front and top of the sample are hidden as well to allow for a view inside the sample. 
  }
  \label{figS2}
\end{figure*}

\begin{figure*}[htbp]
  \includegraphics[width=\textwidth]{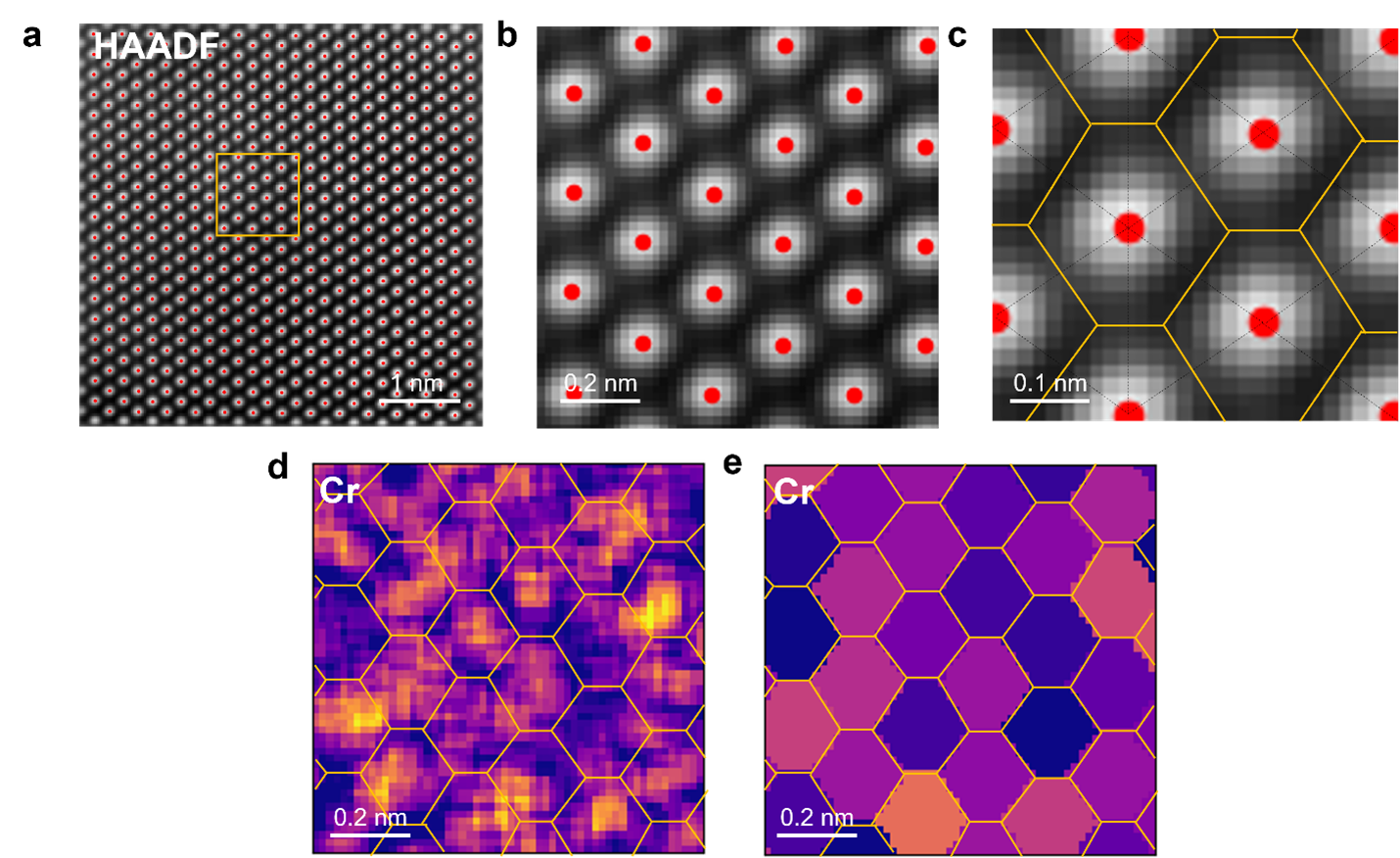}
  \caption{%
  Voronoi cell representation of EDS composition map. 
  \textbf{a} STEM HAADF image recorded in \hkl[110] zone axis. 
  \textbf{b} Magnified STEM HAADF image showing the position (red dot) of atomic columns determined by Gaussian fitting of the column intensity. 
  \textbf{c} Voronoi cells constructed based on the atomic column position. 
  \textbf{d} Voronoi cells overlaid on EDS concentration map of Cr. The (\(5 \times 5\)) binning was applied to the original pixels of EDS map. 
  \textbf{e} EDS composition map generated by averaging the data within each Voronoi cell. 
  }
  \label{figS3}
\end{figure*}

\begin{figure*}[htbp]
  \includegraphics[width=\textwidth]{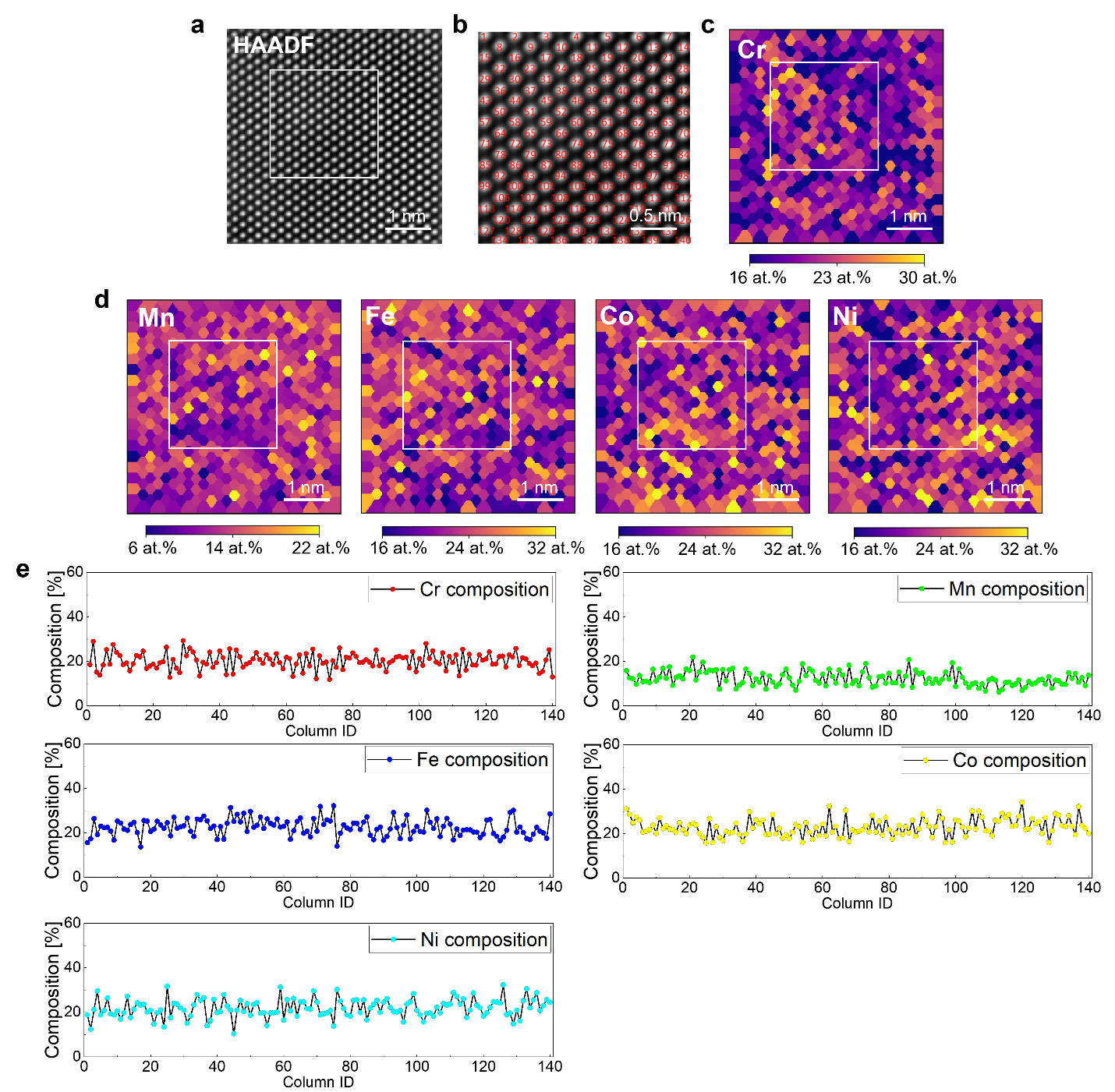}
  \caption{%
  EDS composition maps of dislocation-free region. 
  \textbf{a} STEM HAADF image recorded in \hkl[110] zone axis. 
  \textbf{b} Magnified view of STEM HAADF image (white-lined box in \textbf{a}) showing the atomic columns indexed for composition plots in \textbf{c-e}. EDS composition map of \textbf{c} Cr and \textbf{d} all the other elements. 
  \textbf{e} EDS composition plot of each element within the white-lined box. The column ID is defined in \textbf{b}. Each element shows random fluctuation without noticeable clustering. 
  }
  \label{figS4}
\end{figure*}

\begin{figure*}[htbp]
  \includegraphics[width=\textwidth]{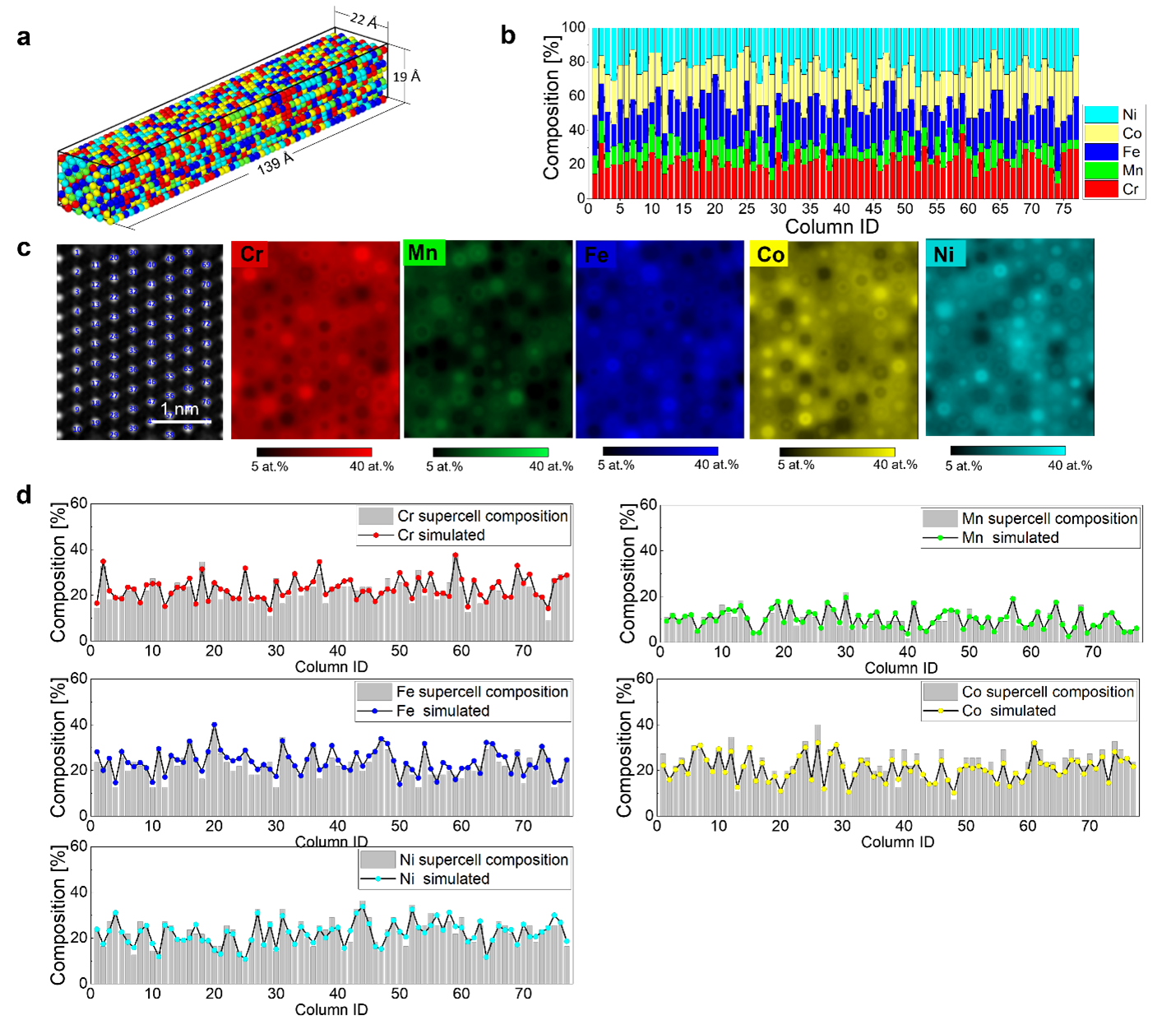}
  \caption{%
  Simulated EDS composition maps of a randomly configured HEA. 
  \textbf{a} Atomic simulation cell. 
  \textbf{b} Plot showing the composition of each atomic column in the simulation cell. 
  \textbf{c} Simulated HAADF image and EDS composition maps using \(\mu\)STEM (version 4.5). In the STEM HAADF image the atomic columns are indexed, starting with position 1 on the top left and ending with position 77 at the lower right, for composition plots in \textbf{d}. 
  \textbf{d} Simulated EDS composition (colored symbol and line) and the composition of simulation cell (gray bar) plotted for the atomic columns defined in \textbf{c}. The two data show almost perfect coincidence. 
  }
  \label{figS5}
\end{figure*}

\begin{figure*}[htbp]
  \includegraphics[width=\textwidth]{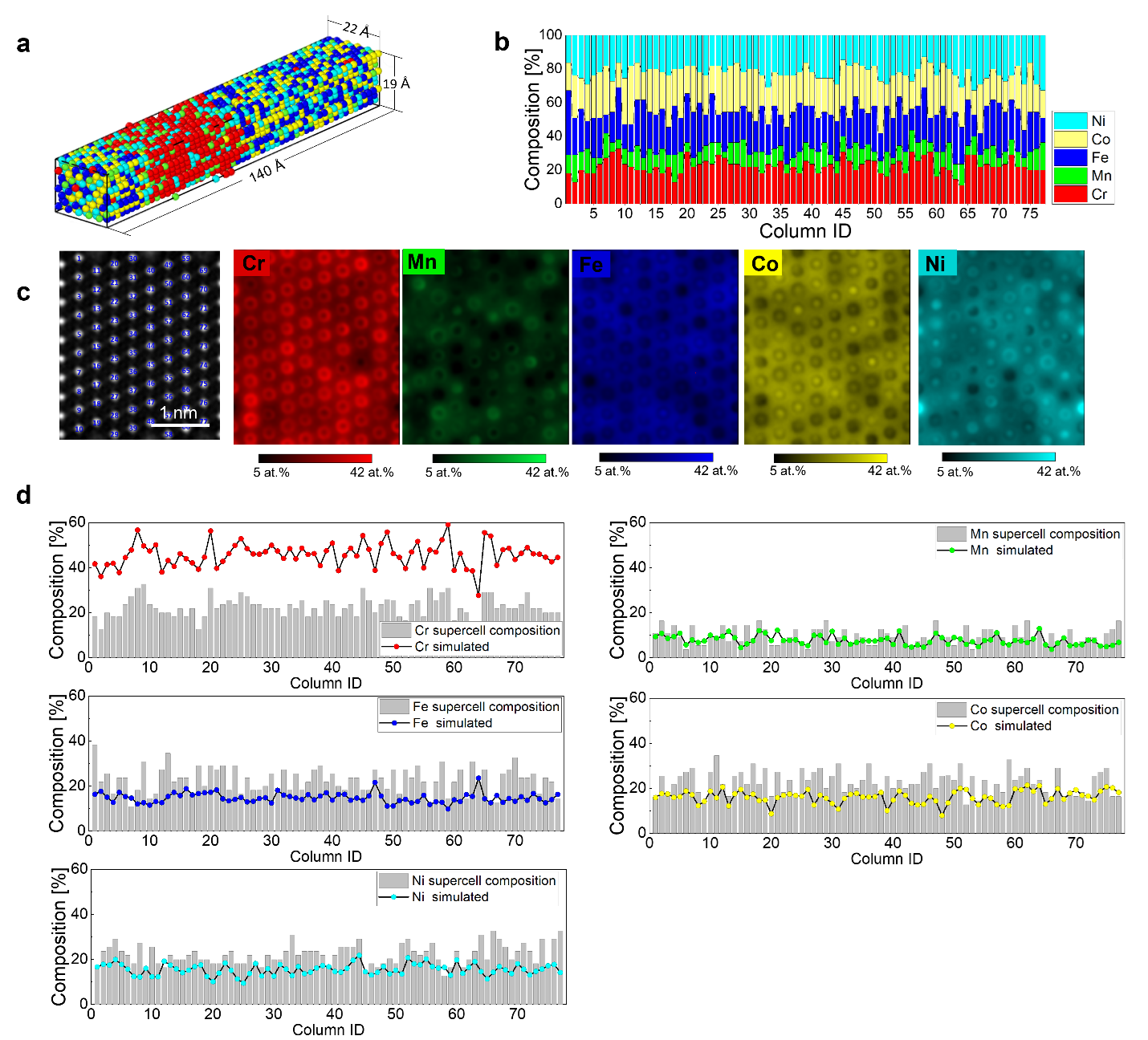}
  \caption{%
  Simulated EDS composition maps of HEA with partially ordered Cr. 
  \textbf{a} Atomic simulation cell. 
  \textbf{b} Plot showing the composition of each atomic column in the simulation cell. Note that the concentration of all elements remained the same as in the randomly configured alloy used for Fig.~\ref{figS5} but only their spatial distribution has been altered. 
  \textbf{c} Simulated HAADF image and EDS composition maps. In the STEM HAADF image the atomic columns are indexed from 1 (top left) to 77 (bottom right) for composition plots in \textbf{d}. 
  \textbf{d} Simulated EDS composition (colored symbol and line) and the composition of simulation cell (gray bar) plotted for the atomic columns defined in c. The composition of the partially ordered element Cr is overestimated by a factor of two whereas that of other elements is slightly underestimated due to the channeling effects on X-ray generation. 
  }
  \label{figS6}
\end{figure*}

\begin{figure*}[htbp]
  \includegraphics[width=\textwidth]{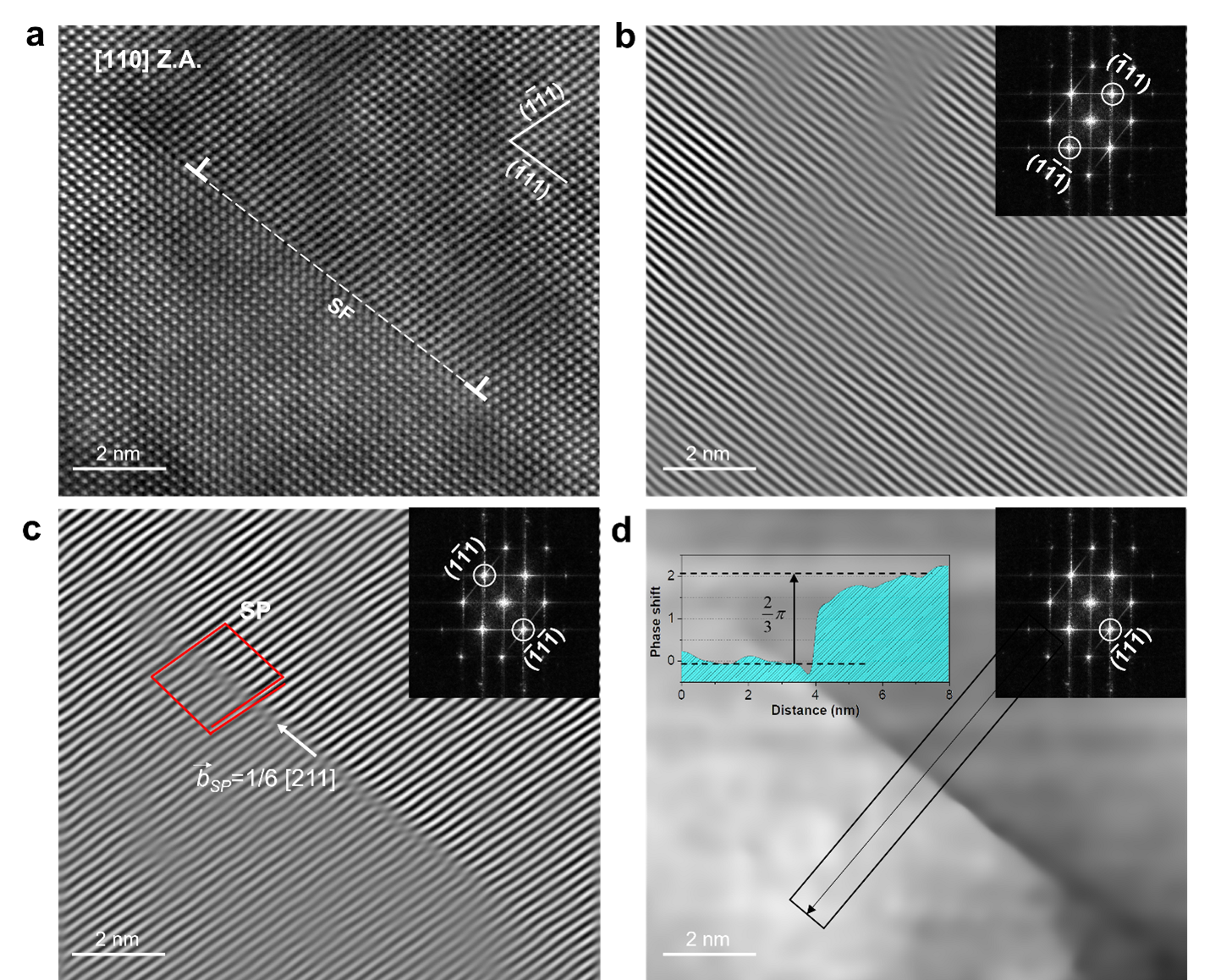}
  \caption{%
  Analysis of the Burgers vector of Shockley partial dislocation. 
  \textbf{a} STEM HAADF image showing an extended dislocation consisted of Shockley partial (SP) dislocations and SF. 
  \textbf{b} Bragg filtered image formed by selecting \hkl(-111) and \hkl(1-1-1) reflections. 
  \textbf{c} Bragg filtered image formed by selecting \hkl(1-11) and \hkl(-11-1) reflections. A Burgers circuit is drawn on SP, showing the Burgers vector of \(a_0/6\hkl[211]\). 
  \textbf{d} Geometric phase image obtained by GPA showing a phase shift of \(2\pi/3\) upon crossing the SF.
  }
  \label{figS7}
\end{figure*}

\begin{figure*}[htbp]
  \includegraphics[width=\textwidth]{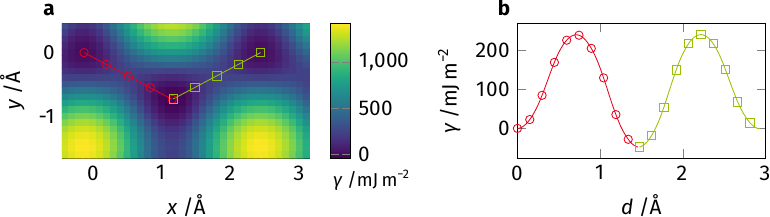}
  \caption{%
  Generalized stacking fault curve of the equimolar Cantor alloy. 
  \textbf{a} GSF energy landscape for the leading (red) and trailing (green) partial dislocations' displacement in the Cantor alloy sample. Averaging is done over the whole glide plane. The local (FCC) and global (HCP) minima configuration are connected via linear displacements to obtain the energy landscape given in \textbf{b}.
  }
  \label{figS8}
\end{figure*}

\begin{figure*}[htbp]
  \includegraphics[width=\textwidth]{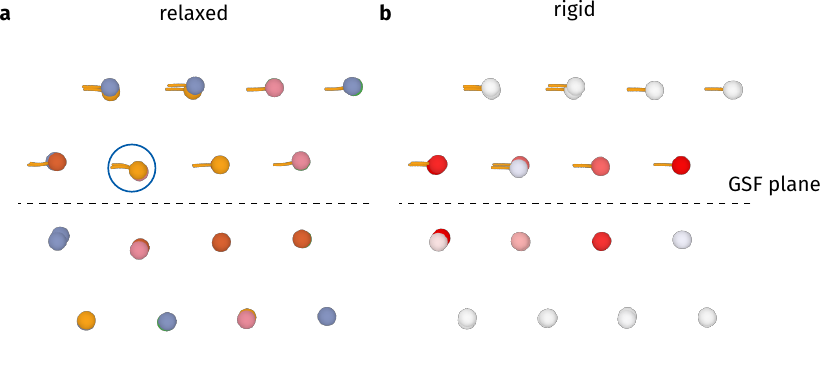}
  \caption{%
  Comparison of the conventional and the rigid relaxation method. Comparison of the two different SF calculation relaxation methods. Both snapshots show the saddle point configuration which corresponds to the unstable SF. Yellow lines show the full atomic trajectory starting from the ideal FCC configuration. 
  \textbf{a} Standard methodology allowing for atomic relaxation normal to the fault plane. An atom with strong structural relaxation is highlighted. The atoms are color coded based on the species and the trajectory lines reveal atomic relaxations. 
  \textbf{b} Proposed method, where the atoms are only allowed to follow rigid body relaxations. The atoms are color coded based on their atomic energy change during this calculation (red corresponds to an energy increase) which is subsequently used to calculate the atomic generalized stacking fault (GSF) energy curves (Fig.~\ref{fig3}a-b). Here, all trajectory lines are exactly parallel.
  }
  \label{figS9}
\end{figure*}

\begin{figure*}[htbp]
  \includegraphics[width=\textwidth]{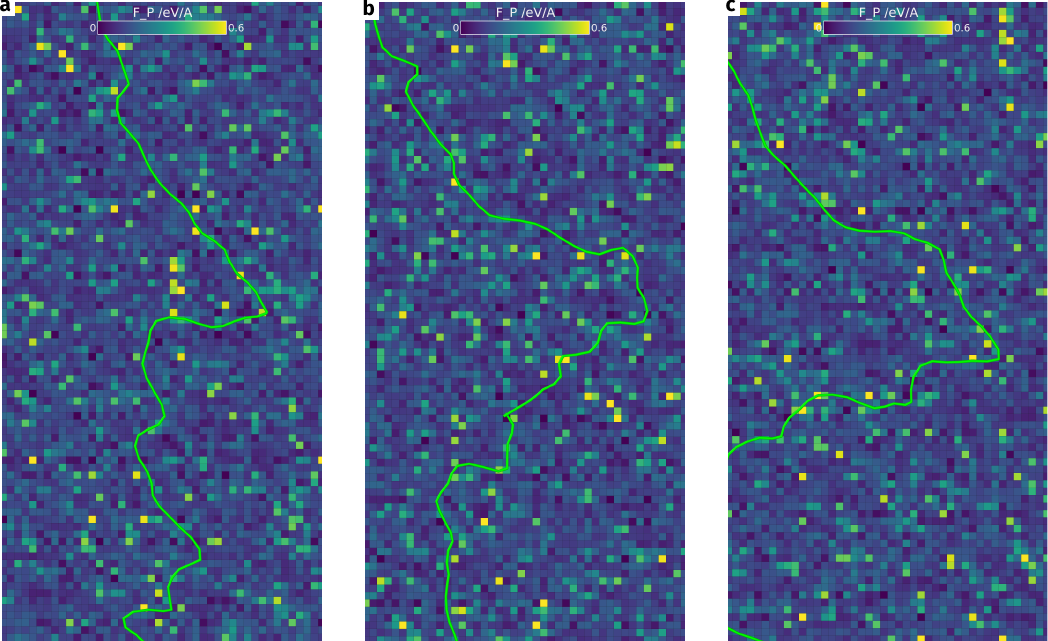}
  \caption{%
  Dislocation pinning of the leading partial in the Cantor alloy. Dislocation lines (green) during glide under constant load in the Cantor alloy superimposed on the local pinning point strength (\Fp{} is denoted as F\_P in the color bar). Here, \Fp{} is averaged in the two lattice planes adjacent to the dislocation glide plane on a \(3 \times 3\)\,\si{\angstrom\squared} grid. The snapshots shown in panels \textbf{a-c} are taken at different times during the simulation. They correspond to \SI{116.04}{\pico\second}, \SI{125.34}{\pico\second}, and \SI{112.65}{\pico\second}, respectively. Note, that each images shows a different section of the dislocation line. 
  }
  \label{figS10}
\end{figure*}

\begin{figure*}[htbp]
  \includegraphics[width=\textwidth]{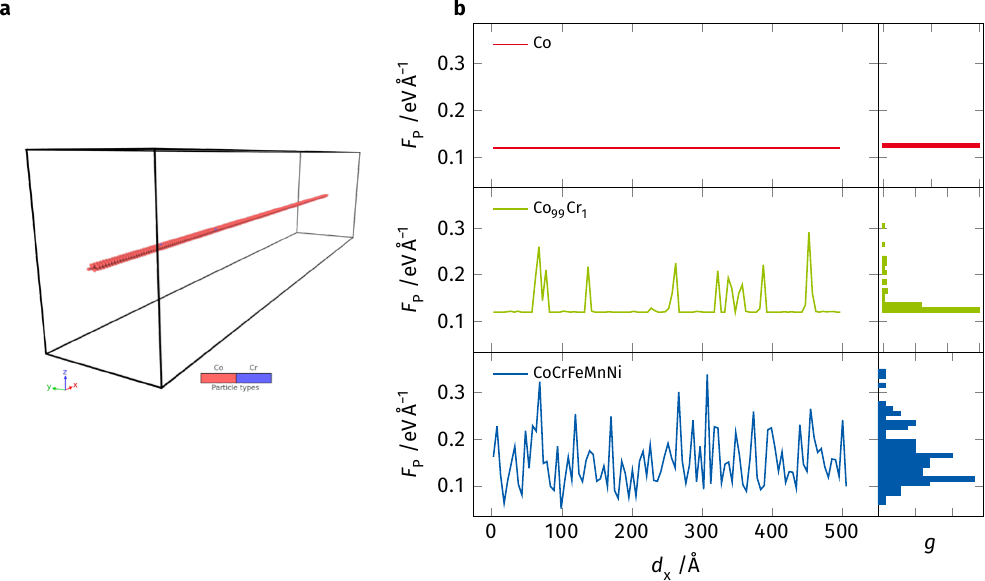}
  \caption{%
  Local friction landscape encountered by a dislocation line as it moves through a pure metal, a dilute solid solution, and a concentrated high-entropy alloy. 
  \textbf{a}Column of atoms above and below the GSF plane measuring 6 atoms across used in the subsequent analysis. The GSF plane (normal to the z-direction) is located between these two rows. Here, the Co\(_{99}\)Cr\(_{1}\) alloy is shown as an example as we expect a strong interaction of Co and Cr atoms based on our results shown in Fig.~\ref{fig4}. 
  \textbf{b} Average \Fp{} values over 100 spatial bins along the x-direction (parallel to the Burgers vector, normal to the dislocation line direction) in Co, Co\(_{99}\)Cr\(_{1}\), and CoCrFeMnNi. On the right the distribution \(g\) of \Fp{} in each sample is shown. In essence, there are no strong pinning points in the pure Co metal, some discrete strong pinning sites in the dilute  Co\(_{99}\)Cr\(_{1}\) solid solution and an abundance of pinning spots in the full HEA. The strong pinning spots in the CoCrFeMnNi alloy do not only show an increase in frequency but also higher peak \Fp{} values. The data confirms the schematic representation given in Fig.~\ref{fig5} of the main manuscript. Note that the \Fp{} data cannot be integrated to derive derive the corresponding energy landscape due to the insufficient spatial resolution.
  }
  \label{figS11}
\end{figure*}

\begin{figure*}[htbp]
  \includegraphics[width=\textwidth]{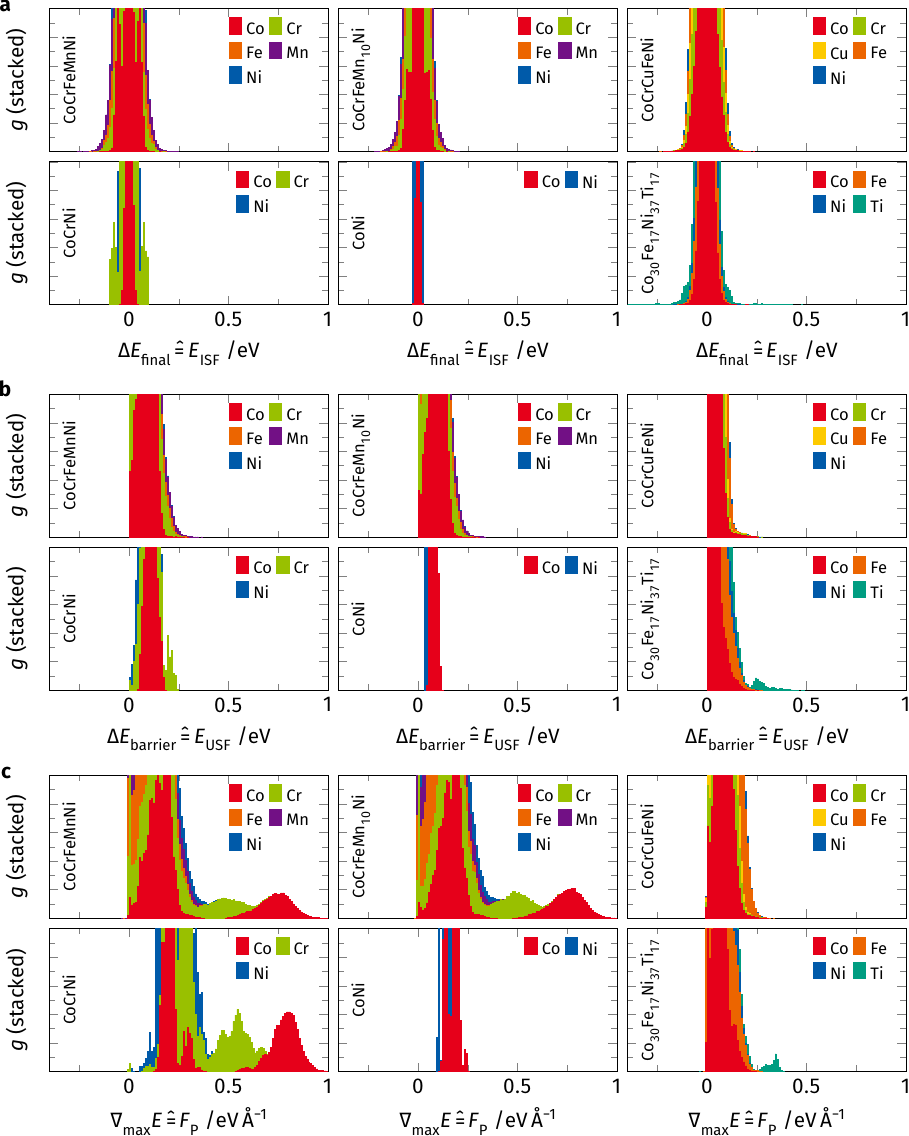}
  \caption{%
  Comparison of the distributions of different atomic descriptors for local dislocation pinning extracted from the GSF curve within different samples. Here we compare: Yhe stable SFE (labeled \(\Delta E_\text{final}\) as it is the energy difference between initial and final state) (\textbf{a}), and the unstable SFE (labeled \(\Delta E_\text{barrier}\) as it is the height of the energy barrier) (\textbf{b}), and \Fp{} (\textbf{c}) for different material systems. All data can be extracted from the per atom energy landscapes calculated for Fig.~\ref{fig3} of the main manuscript. The data representation corresponds to Fig.~\ref{fig4}a.
  }
  \label{figS12}
\end{figure*}

\begin{figure*}[htbp]
  \includegraphics[width=\textwidth]{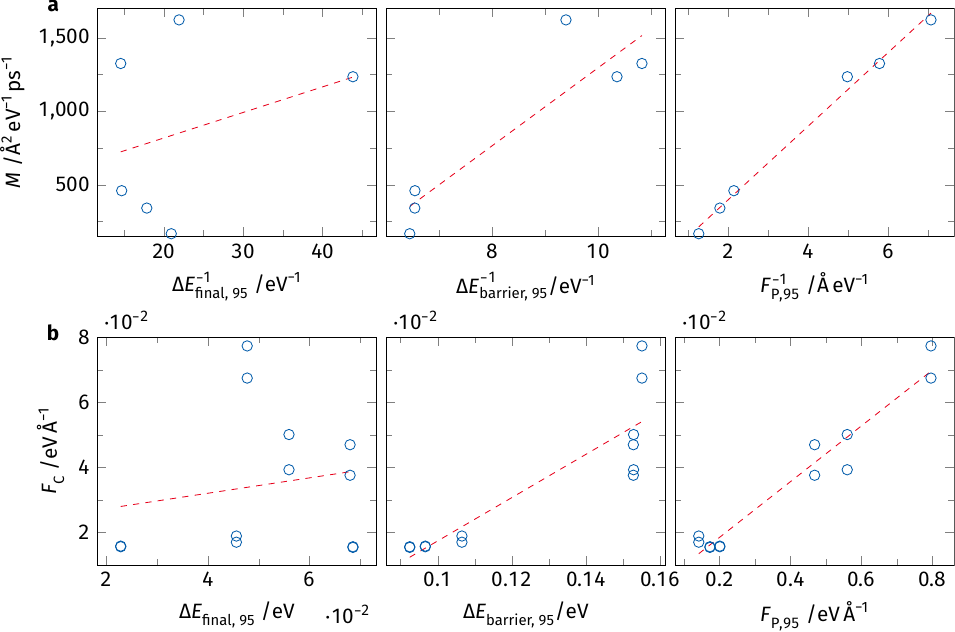}
  \caption{%
  Relation of the critical force and dislocation mobility measured in all samples to the different atomic descriptors. Correlation of the 95th percentile of each descriptor shown in Supplementary Fig.~\ref{figS12} and discussed in the main manuscript with respect to the dislocation mobility \(M\) (\textbf{a}) and the critical force \Fcrit{} (\textbf{b}) required to initiate dislocation glide. The red line corresponds to the best linear fit through the data extracted from the simulations (marks). Best correlation can be seen for the descriptor based on \Fp{}.
  }
  \label{figS13}
\end{figure*}

\cleardoublepage
\section{Supplementary Methods}\label{sec_supp_meth}

Step by step instructions to calculate the per-atom GSF curves
\begin{itemize}
  \item Prerequisites:
  \begin{itemize}
    \item Establish the equilibrium lattice constant of desired composition at \SI{0}{\kelvin}.
  \end{itemize}
  \item Sample preparation:
  \begin{itemize}
    \item {111} plane aligned with one of the cartesian directions (in our case \(z\)-axis)
    \item All atoms on ideal FCC lattice sites with the equilibrium lattice spacing
    \item Alloy composition and atomic ordering as desired
  \end{itemize}
  \item Simulation setup:
  \begin{itemize}
    \item Define 2 groups of atoms one above and one below the GSF \hkl{111} plane
    \item Periodic boundary conditions in the GSF plane, open boundaries normal to the GSF plane
  \end{itemize}
  \item Simulation:
  \begin{itemize}
    \item Scan atoms of group 1 across group 2, i.e., displace them in \(x\)- and \(y\)-direction (assuming the \(z\)-direction is \hkl{111} aligned). The scan direction / area depends on the Burges vector on the dislocation of interest.
    \item Relax the system at each displacement grid point and calculate the energy. For relaxation, the forces in \(x\)- and \(y\)- direction (in the GSF plane) are set to 0, while the forces in \(z\)-direction (normal to the GSF plane) are averaged across all atoms in each group, respectively. This results in a rigid body displacement of the two regions (above and below the GSF plane) against each other.
    \item Store the energy after each minimization (at each grid point) for each atom.
  \end{itemize}
  \item Post processing:
  \begin{itemize}
    \item Collect the displacement-energy maps for each atom in the GSF plane
    \item Apply interpolation using, e.g., bicubic splining
    \item Find the local minima (cf. Fig.~\ref{fig3}) corresponding to the (meta-)stable lattice positions and connect them using straight lines.
    \item Extract maximum gradients along these profiles
  \end{itemize}
\end{itemize}

\end{document}